\documentclass[11pt, a4paper]{scrartcl}

\usepackage[utf8]{inputenc}
\usepackage[english]{babel}
\usepackage{hyperref}
\usepackage{amsmath,amssymb,amsfonts,mathrsfs, amsthm}
\usepackage{mathtools}
\usepackage{amsthm}
\usepackage{dsfont}
\usepackage{bm}
\usepackage{colortbl}
\usepackage{fullpage}
\usepackage[ruled, linesnumbered]{algorithm2e}
\usepackage{multirow,booktabs,bigdelim}
\usepackage{caption}
\usepackage{comment}
\usepackage{graphicx}
\usepackage{placeins}
\usepackage{natbib}
\usepackage{siunitx}
\usepackage{url}
\usepackage{paralist}
\usepackage{enumitem}
\usepackage{layouts}
\usepackage[rgb]{xcolor}
\usepackage{scalerel}
\usepackage{tikz}
\usepackage{tkz-graph}
\usepackage{authblk}
\usetikzlibrary{shapes.geometric}
\usetikzlibrary{backgrounds}
\usetikzlibrary{arrows.meta}
\usepackage[framemethod=TikZ]{mdframed}

\definecolor{col1}{RGB}{88,140,126}
\definecolor{col2}{RGB}{242,227,148}
\definecolor{col3}{RGB}{242,174,114}
\definecolor{col4}{RGB}{217,100,89}
\definecolor{col5}{RGB}{140,70,70}
\colorlet{lightgray}{black!15}

\newcommand{\R}{\mathbb{R}}
\newcommand{\N}{\mathbb{N}}

\DeclareMathOperator*{\argmin}{arg\,min}
\DeclareMathOperator*{\argmax}{arg\,max}

\DeclarePairedDelimiterX{\norm}[1]{\lVert}{\rVert}{#1}
\DeclarePairedDelimiterX{\abs}[1]{\lvert}{\rvert}{#1}

\renewcommand{\epsilon}{\varepsilon}

\renewcommand{\P}{\mathbb{P}}
\newcommand{\E}{\mathbb{E}}

\newcommand{\independent}{\perp\!\!\!\perp}

\newcommand{\independentG}{\perp\!\!\!\perp_\mathcal{G}}
\newcommand{\nindependentG}{\not\!\perp\!\!\!\perp_\mathcal{G}}
\newcommand{\var}{\operatorname{Var}}
\newcommand{\cov}{\operatorname{Cov}}
\newcommand{\vI}{\operatorname{Id}}

\newcommand{\vE}{\mathbf{E}}

\newcommand{\vX}{\mathbf{X}}
\newcommand{\vY}{\mathbf{Y}}

\newcommand{\PA}{\operatorname{PA}}
\newcommand{\CH}{\operatorname{CH}}
\newcommand{\tCH}{\scaleto{\operatorname{CH}}{3.5pt}}
\newcommand{\tPA}{\scaleto{\operatorname{PA}}{3.5pt}}
\newcommand{\DE}{\operatorname{DE}}
\newcommand{\SB}{\operatorname{SB}_I}
\newcommand{\NSB}{\operatorname{NSB}_I}
\newcommand{\MB}{\operatorname{MB}}

\newcommand{\stabscore}{\textbf{s}_{\operatorname{stab}}}
\newcommand{\predscore}{\textbf{s}_{\operatorname{pred}}}

\numberwithin{equation}{section}

\theoremstyle{plain}
\newtheorem{theorem}{Theorem}[section]
\newtheorem{example}[theorem]{Example}
\newtheorem{corollary}[theorem]{Corollary}
\newtheorem{lemma}[theorem]{Lemma}
\newtheorem{proposition}[theorem]{Proposition}

\theoremstyle{definition}
\newtheorem{definition}[theorem]{Definition}
\newtheorem{setting}{Setting}

\theoremstyle{remark}

\title{Stabilizing Variable Selection and Regression}

\author[1]{Niklas Pfister}
\author[2]{Evan G. Williams}
\author[3]{Jonas Peters}
\author[4]{Ruedi Aebersold}
\author[5]{Peter B\"uhlmann}
\affil[1,5]{Seminar for Statistics, ETH Z\"urich, Switzerland}
\affil[2,4]{Department of Biology, ETH Z\"urich, Switzerland}
\affil[3]{Department of Mathematics, University of Copenhagen, Denmark}

\begin{document}
\maketitle

\begin{abstract}
  We consider regression in which one predicts a response $Y$ with a
  set of predictors $X$ across different experiments or
  environments. This is a common setup in many data-driven scientific
  fields and we argue that statistical inference can benefit from an
  analysis that takes into account the distributional changes across
  environments. In particular, it is useful to distinguish between
  stable and unstable predictors, i.e., predictors which have a fixed
  or a changing functional dependence on the response,
  respectively. We introduce stabilized regression which explicitly
  enforces stability and thus improves generalization performance to
  previously unseen environments. Our work is motivated by an
  application in systems biology. Using multiomic data, we demonstrate
  how hypothesis generation about gene function can benefit from
  stabilized regression. We believe that a similar line of arguments
  for exploiting heterogeneity in data can be powerful for many other
  applications as well. We draw a theoretical connection between
  multi-environment regression and causal models, which allows to
  graphically characterize stable versus unstable functional
  dependence on the response. Formally, we introduce the notion of a
  stable blanket which is a subset of the predictors that lies between
  the direct causal predictors and the Markov blanket. We prove that
  this set is optimal in the sense that a regression based on these
  predictors minimizes the mean squared prediction error given that
  the resulting regression generalizes to unseen new environments.
\end{abstract}

\section{Introduction}

Statistical models usually describe the observational distribution of
a data generating process. In many applied problems this data
generating process may change over time or across experiments. In such
settings, it is useful to get a mechanistic understanding of the
underlying changes in the system; both to understand which parts of a
system cause certain outcomes and to make reliable predictions under
previously unseen conditions. One approach to rigorously model such
changes are causal models \citep{Pearl2009, Imbens2015} which allow
for changes in the data generating process via the notion of
interventions. As demonstrated in Section~\ref{sec:SP_vars}, this
framework can be related to multi-environment regression, hence
creating a link between the two areas of study: (i) learning a
regression which performs well under unseen intervention settings and
(ii) selecting variables based on their behavior under different
observed interventions. Although we use a causal framework for
formulation, we do not necessarily address the ambitious task of
inferring causality but rather aim for a notion of stability and
invariance.  The goal of this paper is to analyze the connection
between (i) and (ii) and use it to develop a methodological framework
for inference.

This study is motivated by an application in systems biology in which
one performs an exploratory analysis to discover the impact of genetic
and environmental variants on known metabolic pathways and phenotypes
\citep[work in progress]{roy2019}.  More specifically, we consider
multiomic data from the transcriptome and proteome of a mouse
population of 57 different inbred strains that was split into two
groups, fed either with a low fat diet or a high fat diet. Liver
tissue from these cohorts was then collected at multiple timepoints
across their natural lifespans, providing diet as an independent
biological (environment) variable. Based on these data, the target of
interest is to associate gene expression of mRNAs and proteins in
central metabolic pathways and using the independent biological
variables to infer causality. This provides two avenues of hypotheses
generation: (1) identifying pathway-associated genes which are not in
the canonical lists, and (2) determining which genes are (causally)
upstream and driving pathway activity across the population as a
function of diet.

\subsection{Stabilized regression}

\begin{figure}
  \centering
  \resizebox{\textwidth}{!}{
    \includegraphics{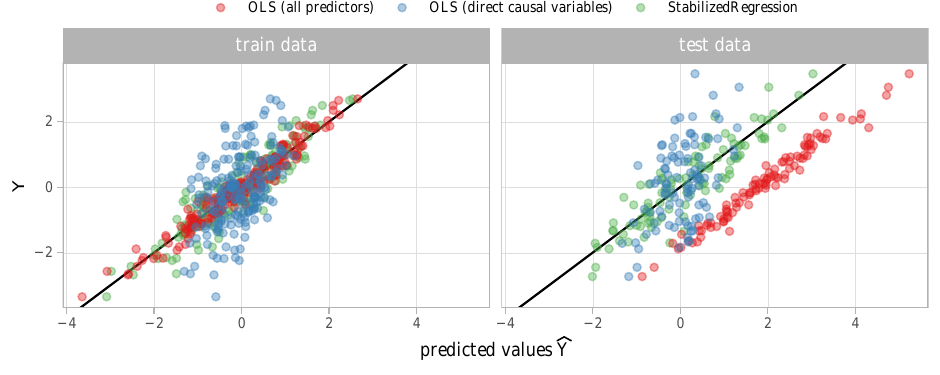}}
  \caption{Illustrative example of three linear regression procedures
    applied to data generated according to
    Example~\ref{ex:toy_example} with two training and one testing
    environment. A good fit means that the dots are close to the
    identity line (given in black). Linear regression based on all
    predictors (red) leads to biased results on the testing
    environment, while a linear regression based only on direct causal
    variables of the response (blue) leads to unbiased estimation but
    with higher variance in both the testing and training
    environments. Stabilized regression (green) aims for the best fit
    which is also unbiased in the unobserved testing environment.}
  \label{fig:illustrative}
\end{figure}

Consider the following multi-environment regression setting; let
$X=(X^1,\dots,X^d)\in\bm{\mathcal{X}}$ be a vector of predictor
variables and $Y\in\R$ a response variable, both of which are observed
in different (perturbation) environments $e \in \mathcal{E}$.  We
assume that in each environment $e\in\mathcal{E}$, the variables
$(Y_e, X_e)$ have joint distribution $P_e$. Assume further that we
only observe data from a subset of the environments
$\mathcal{E}^{\text{obs}}\subseteq\mathcal{E}$. For each observed
environment there are i.i.d.\ data, yielding $n$ observations across
all observed environments. The data can thus be represented by an
($n\times d$)-matrix $\vX$, an ($n\times 1$)-vector $\vY$ and an
($n\times 1$)-vector $\vE$ indicating which experiment the data point
comes from. The special case of an underlying linear model is shown in
Figure~\ref{fig:illustrative} (left: observed training data, right:
unobserved test data) with data generated according to a Gaussian
linear model consisting of shift environments
(Example~\ref{ex:toy_example}). The data have been fitted on the
training environment using linear regression on all variables (red)
and on only the direct causal variables of the response (blue) --
which might be unknown in practice, of course. Since the underlying
data generation process changes across settings, the regression based
on all predictors leads to a biased prediction in the unobserved test
environment, while the regression based only on the direct causal
variables allows to generalize to these settings. At the same time the
fit of the model based solely on the direct causal variables has
higher variance on both training and test environments compared with
the regression based on all predictors. The method we describe in this
paper attempts (without knowing the underlying model) to be able to
generalize to unseen settings without bias, while at the same time
minimizing the prediction error. In Figure~\ref{fig:illustrative}, we
show the result of the proposed method in green.

Assuming an underlying causal structure, there is a key relation
between a regression that is able to generalize and the variables that
can be included into that regression. Details on this connection are
given in Section~\ref{sec:SP_vars}. Using it, we can get a causal
understanding of the underlying mechanism by looking at which sets of
predictors lead to models that generalize and which do not.

\begin{figure}
  \centering
  \resizebox{\textwidth}{!}{
    \includegraphics{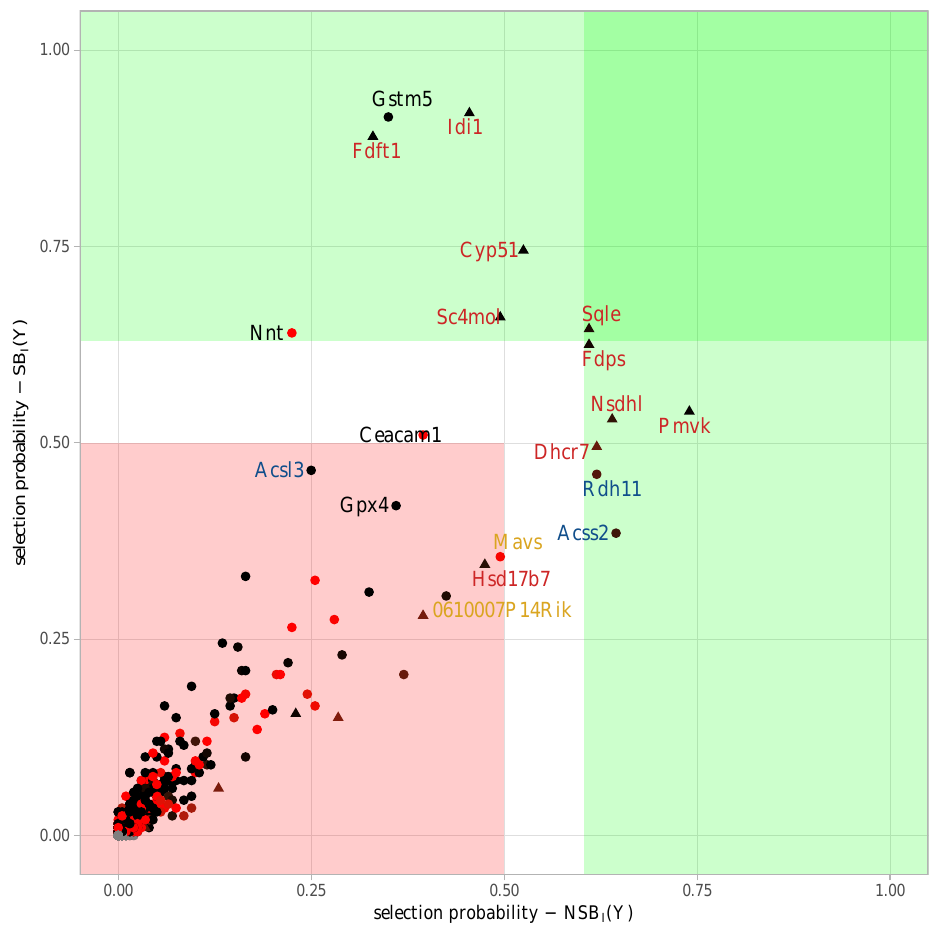}}
  \caption{Stabilized regression (SR) applied to the \emph{Cholesterol
      Biosynthesis} pathway (CB). The data set consists of protein
    expression levels ($n=315$) measured for $d=3939$ genes, $16$ of
    which are known to belong to CB (red gene names). We take protein
    expression levels of one known CB gene (Hmgcs1) as response
    $Y$. On the x- and y-axis we plot subsampling-based selection
    probabilities for two SR based variable selection procedures;
    y-axis: stable genes $\SB(Y)$ and x-axis: non-stable genes
    $\NSB(Y)$ (The precise definitions can be found in
    Section~\ref{sec:SP_vars}.)  Many significant genes (green area)
    are canonical CB genes (red label) or part of an adjacent pathway
    (blue label). Annotated genes with a semi-evident relationship
    have yellow labels and with no clear relation black labels. The
    color coding of the nodes (interpolating between red and black)
    corresponds to the fraction of times the sign of the regression
    coefficient was negative/positive (red: negative sign, black:
    positive sign, grey: never selected).}
  \label{fig:pathway_example}
\end{figure}

In the gene function discovery application
\citep[][e.g.,]{francesconi2008, dutkowski2013}, one wants to find
novel gene relationships that can be associated to known
pathways. Furthermore, one is also interested in understanding how a
gene functions within a pathway, for example in the mouse data set
mentioned above, whether it is active in all mice, whether it changes
over time or whether its function depends on diet. Often such question
can be answered by understanding whether a functional dependence
remains fixed or changes depending on some exogenous environment
variable. For an illustration of this problem based on the mouse data
set, consider Figure~\ref{fig:pathway_example}.
There, we consider protein expression levels of $3939$ genes (based on
$n=315$ observations) and try to find functionally related genes to a
known cholesterol biosynthesis gene (Hmgcs1). To do this, we set the
response $Y$ to be the protein expression levels of Hmgcs1 and then
apply stabilized regression together with stability selection. The
exact procedure is described in Section~\ref{sec:application_bio}. In
Figure~\ref{fig:pathway_example}, we plot the selection probabilities
of genes (large probabilities imply we are certain about the finding)
which either have an \emph{unstable} or a \emph{stable} functional
relationship with $Y$ across diets on the x-axis and y-axis,
respectively. The genes have been annotated according to their
relationship to the cholesterol biosynthesis pathway from the Reactome
Pathway Knowledgebase \citep{Reactome}, which consists of $25$ known
canonical pathway genes of which $16$ have been measured (including
Hmgcs1). The result shows that stabilized regression is able to
recover many relevant genes and also allows to group findings into
stable and unstable relationships. Details about the labeled genes and
their relation with the cholesterol pathway is given in
Appendix~\ref{sec:bio_annotations}.

To achieve these goals, we propose a stabilizing procedure that can be
combined with an arbitrary regression technique for each environment
$e\in\mathcal{E}^{\text{obs}}$ individually. More specifically, for
any subset $S\subseteq\{1,\dots, d\}$, let $\hat{f}^{S}$ be a
regression estimate as a function of the predictors $X^{S}$. We then
define the \emph{stabilized regression} estimator to be a weighted
average of the following form
\begin{equation}
  \label{eq:stabilizedregression_estimator}
  \hat{f}_{\operatorname{SR}}(X)\coloneqq\sum_{S\subseteq\{1,\dots,
    d\}}\widehat{w}_S\cdot \hat{f}^{S}(X^{S}),
\end{equation}
where $\widehat{w}_S$ are normalized weights, i.e.,
$\sum_{S}\widehat{w}_S=1$. This type of model averaging appears often
in the literature and we discuss related approaches in
Section~\ref{sec:related_work}. Commonly, the weights are chosen to
optimize the predictive performance of the averaged model (e.g., by
considering the residual sum of squares or various information
criteria).  We propose, however, that large weights should be given to
models which are both stable and predictive.  Here, stability means
that the models do not vary much between the different environments.
We provide a formal definition in
Section~\ref{sec:generalizing_regression}, but other choices are
possible, too, and may be of particular interest for complex data
structures, such as dynamical data \citep{Pfister2019causalkinetix}.

\subsection{Related work}\label{sec:related_work}

Performing prediction in new unobserved perturbed or changed
environments is of huge importance in many applied areas and has been
termed transfer learning or domain adaption in the machine learning
and statistics community. While there are many different types of
modeling frameworks for this problem, one very well-established idea
is to use causal models \citep{Pearl2009} and formalize the changes
across environments by the notion of interventions. The key idea
behind this approach is that causal models offer an intuitive way of
modeling the conditional distribution of the response $Y$ given its
predictors $X$. More specifically, a causal model implies invariance
of the conditional distribution under certain conditions, which can be
used to perform prediction in unseen environments. This is a
fundamental concept in causality and has been referred to as
invariance, autonomy or modularity \citep{Wright1921, Haavelmo1944,
  Aldrich1989, Hoover1990, Imbens2015, richardson2013single}. The
invariance principle can be used to learn parts of a causal models
from data and hence give a causal interpretation to some of the
variables. This can be done by turning the invariance assumption
around and inferring a causal model by finding models which remain
invariant. Using this idea to find direct causes of a response has
been done in \citet{Peters2016jrssb, Pfister2018jasa} and
\citet{heinze2018invariant}. On the other hand, one can also use the
invariance principle to improve prediction on unseen
environments. Several existing methods learn models that explicitly
enforce this assumption in order to generalize to new settings, as for
example, \citet{icml2012, zhang2013domain, Rojas2016} and
\citet{heinze2017conditional}. Others have tried to weaken the
invariance assumption by only penalizing the non-invariance and hence
trading-off generalization with in-sample prediction performance
\citep[e.g.,][]{pan2010domain, ganin2016domain,
  rothenhausler2018anchor}. A general discussion, about the relation
of invariance and causality is given by
\citet{buhlmann2018invariance}. Our proposed framework incorporates
the idea of using invariance in order to improve generalization, while
at the same time aiming for a causal interpretation of the resulting
variable selection.

From an algorithmic point of view, our proposed method is related to
several averaging techniques from the literature. Averaging is a
common regularization principle throughout statistics with many
different types of applications in regression and variable
selection. The idea of aggregating over several models is for example
done in the generalized ensemble method due to \citet{perrone1992},
which gives explicit equations for optimal weights in terms of
prediction MSE. Similar ideas, also exist in the Bayesian community,
termed Bayesian model averaging (BMA)
\citep{hoeting1999bayesian}. There, models are aggregated by
optimizing the posterior approximation based either on the Bayesian
information criterion (BIC) \citep{schwarz1978estimating} or on the
Akaike information criterion (AIC) (leading to the so-called Akaike
weights due to \citet{burnham1998practical}). Our stabilized
regression estimator in \eqref{eq:stabilizedregression_estimator}
averages over all subsets of predictors which is similar to how for
example random forests \citep{breiman2001random} are
constructed. Other related approach based on resampling subsets of
predictors are due to \citet{wang2011random} and
\citet{cannings2017}. Our method is, however, unique in combining this
type of averaging with environment-wise stability or invariance.

Finally, the notion of stability has been widely used in several
related contexts in statistics. As pointed out by for example
\citet{Yu2013} and \citet{Yu2019}, reproducible research relies on the
statistical inference being stable across repetitions of the same
procedure. This idea also underlies well-established resampling
schemes such as bagging by \citet{breiman1996bagging} and stability
selection by \citet{meinshausen2010stability}.

\subsection{Contributions} 
We introduce a novel regression framework based on averaging that
allows to incorporate environment-wise stability into arbitrary
regression procedures. Under mild model assumptions, our resulting
regression estimates are shown to generalize to novel environmental
conditions. The usefulness of our procedure is demonstrated for an
application about gene detection from systems biology. For this
application, besides using our novel stabilized regression, we propose
an additional graphical tool which allows to visualize which genes are
related to a response variable and whether this relationship is stable
or unstable across environments. We believe this can aid practitioners
to explore novel biological hypotheses. Finally, we introduce a
theoretical framework for multi-environment regression and prove
several results which relate it to structural causal models. Based on
this correspondence we introduce the stable blanket $\SB(Y)$, a subset
of the Markov blanket, and discuss how this might help interpreting
the output of different variable selection techniques. Our procedure
will be made available as an easy-to-use R-package.

\subsection{Outline}

In Section~\ref{sec:generalizing_regression}, we define our formal
target of inference and describe the multi-environment regression
setting. Then, in Section~\ref{sec:SP_vars}, we propose a causal
model framework and prove theoretical results relating the causal
model perspective and multi-environment regression. Moreover, we
introduce the concept of a stable blanket and discuss, how this allows
us to interpret different variable selection techniques. This section
can be skipped by the practical-minded reader. Our proposed algorithm is
presented in Section~\ref{sec:meth_details}, in which we also give
details about practical issues in the implementation. In
Section~\ref{sec:numerical_sim}, we benchmark our method with commonly
employed techniques based on two simulation experiments. Finally, in
Section~\ref{sec:application_bio}, we discuss the biological pathway
analysis application in detail and explain how to construct
visualizations as in Figure~\ref{fig:pathway_example}.

\section{Multi-environment
  regression}\label{sec:generalizing_regression}

Stabilized regression can be seen as a multi-environment regression
technique for domain adaptation or transfer learning. The following
summarizes the technical details of our multi-environment setup.
\begin{setting}[multi-environment regression]
  Let $\bm{\mathcal{X}}=\mathcal{X}^1\times\cdots\times\mathcal{X}^d$
  be a $d$-dimensional product of measurable spaces, let
  $X=(X^1,\dots,X^d)\in\bm{\mathcal{X}}$ be a random vector of
  predictor variables, let $Y\in\R$ be a random response variable and
  let $\mathcal{E}$ be a collection of perturbation environments such
  that for each environment $e\in\mathcal{E}$ the variables
  $(Y_e, X_e)$ have joint distribution $P_e$. We assume that the
  distributions $P_e$ are absolutely continuous with respect to a
  product measure which factorizes. Assume that we only observe data
  from a subset of the environments
  $\mathcal{E}^{\text{obs}}\subseteq\mathcal{E}$.
\end{setting}
Given this setting, our goal is to make predictions on a potentially
unseen environment $e\in\mathcal{E}$. For this to be meaningful, some
assumption on the type of perturbations in $\mathcal{E}$ is
required. Motivated by previous work in causality
\citep[e.g.,][]{Peters2016jrssb}, we assume that that there exists a
subset $S\subseteq\{1,\dots,d\}$ such that for all environments
$e,h\in\mathcal{E}$ and all $x\in\bm{\mathcal{X}}$ it holds that
\begin{equation}
  \label{eq:invariant_mechanism_assumption}
  \E(Y_e\,\vert\, X_e^{S}=x^S)=\E(Y_h\,\vert\, X_h^{S}=x^S).
\end{equation}
As we point out in Section~\ref{sec:SP_vars}, this assumption can be
related to an underlying causal model. In that case, condition
\eqref{eq:invariant_mechanism_assumption} coincides with parts of the
causal system being fixed, which is a fundamental concept referred to
as invariance, autonomy or modularity.

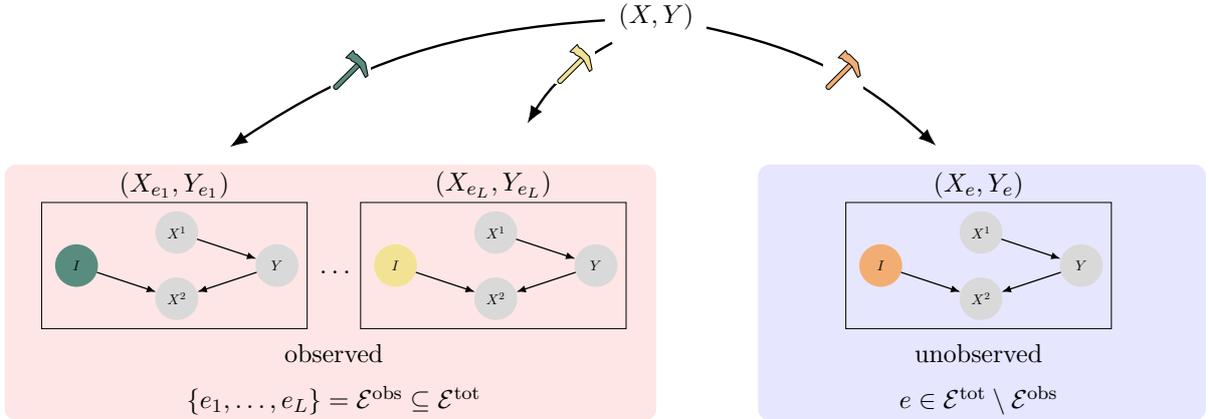
\begin{figure}[ht]
  \centering
  \resizebox{\textwidth}{!}{
    \begin{tikzpicture}[scale=1]
      % unobserved and observed background
      \draw[fill=red!10, draw=none, rounded corners=5] (-9.6,0.3)
      rectangle (0,-3.5);
      \draw[fill=blue!10, draw=none, rounded corners=5] (1.5,-3.5) rectangle (8.1,0.3);
      % Central node
      \node[circle] (S) at (0,2.5){$(X,Y)$};
      % Graph nodes
      \tikzstyle{VertexStyle} = [shape = circle, minimum width = 3em]
      \Vertex[Math,L={(X_{e_1},Y_{e_1})},x=-7.1,y=0]{S1}
      \Vertex[Math,L={(X_{e_L},Y_{e_L})},x=-2.4,y=0]{S2}
      \Vertex[Math,L=\dots,x=-4.7,y=-1.3]{dots}
      \Vertex[x=-4.75,y=-2.5]{observed}
      \Vertex[x=4.75,y=-2.5]{unobserved}
      \node (env1) at (-4.75,-3.2) {$\{e_1,\dots,e_L\}=\mathcal{E}^{\text{obs}}\subseteq\mathcal{E}$};
      \node (env1) at (4.75,-3.2) {$e\in\mathcal{E}\setminus\mathcal{E}^{\text{obs}}$};
      \Vertex[Math,L={(X_e,Y_e)},x=4.75,y=0]{SL1}
      \tikzstyle{VertexStyle} = [shape = square, minimum width = 3em,
      draw]
      % Graph edges
      \tikzstyle{EdgeStyle} = [-Latex, line width=1, bend right=15]
      \Edge(S)(S1)
      \Edge(S)(S2)
      \tikzstyle{EdgeStyle} = [-Latex, line width=1, bend left=15]
      \Edge(S)(SL1)
      % data nodes
      \node (pic1) at (-4.5, 1.7)
      {\includegraphics[width=1.5em]{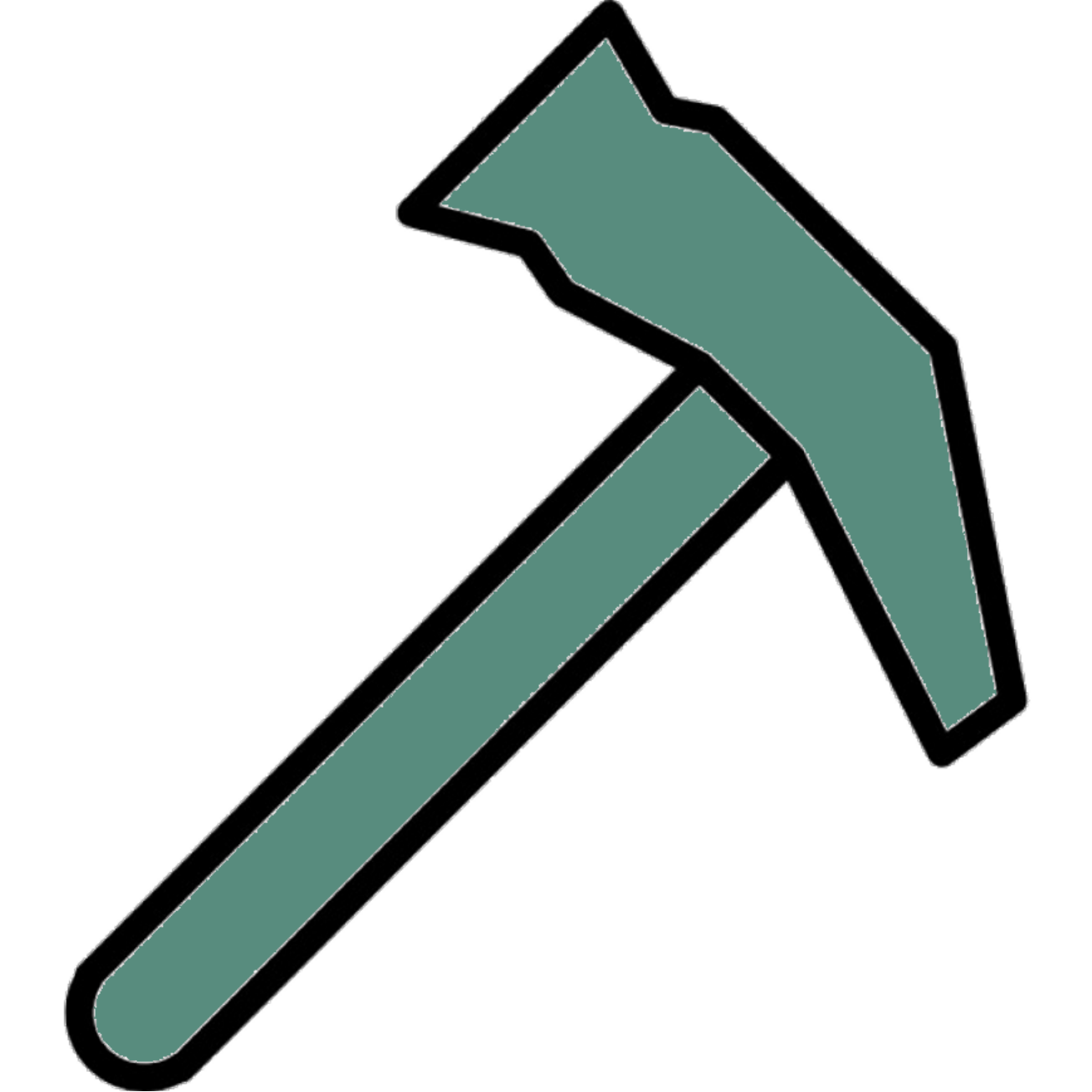}};
      \node (pic2) at (-1.2, 1.8)
      {\includegraphics[width=1.5em]{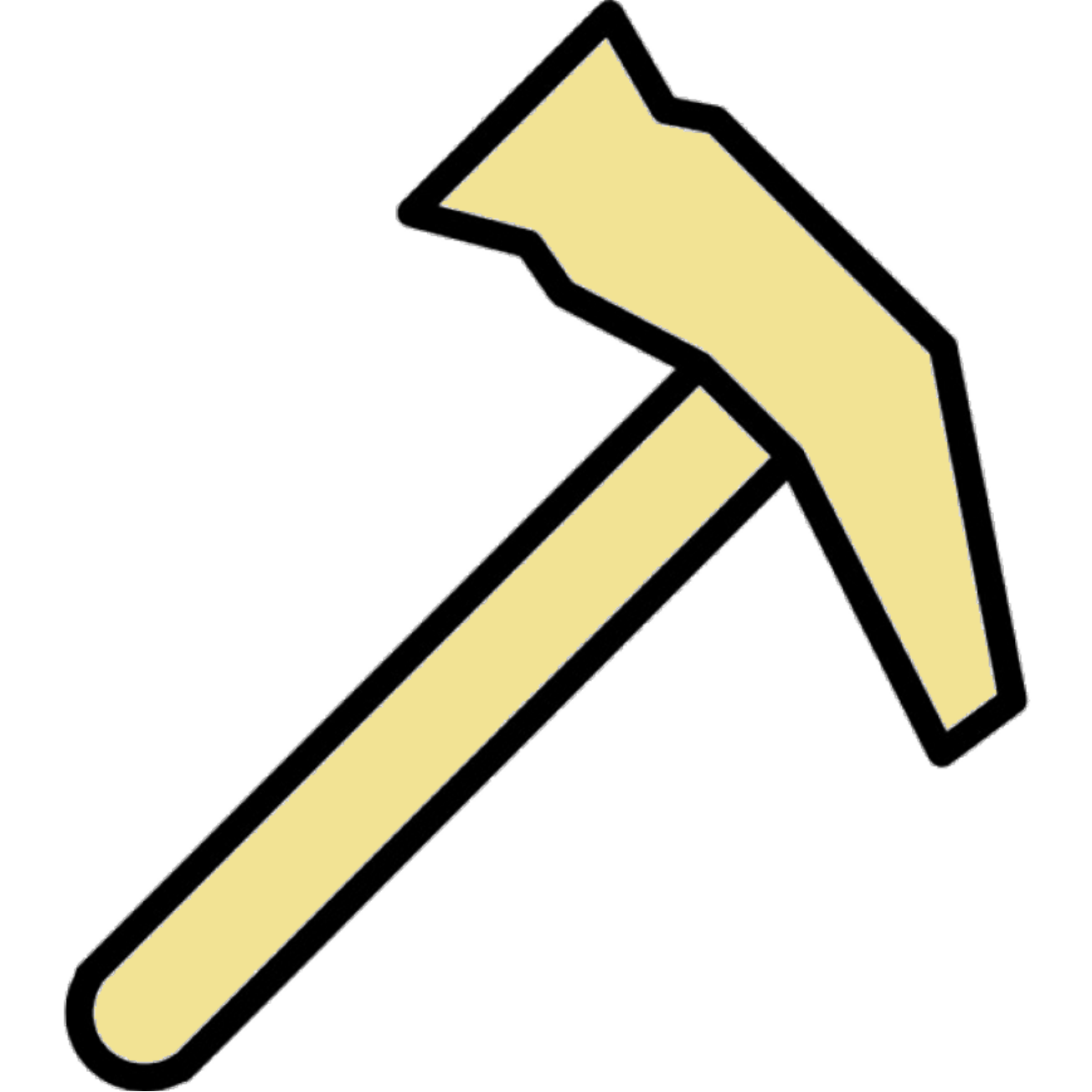}};
      \node (pic3) at (2.75, 1.7)
      {\includegraphics[width=1.5em]{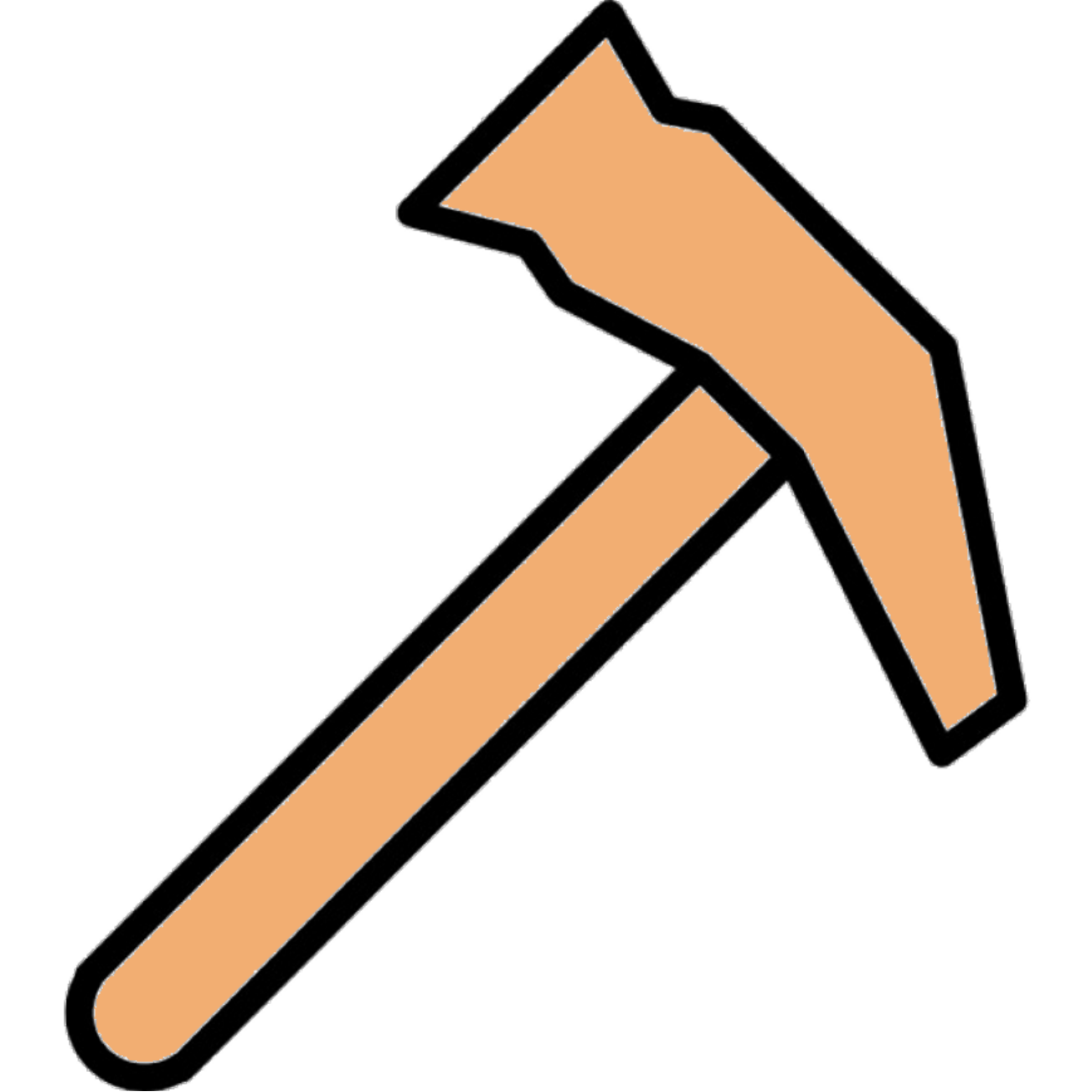}};
      \node[rectangle, draw] (data1) at (-7.1, -1.2) {%
        \begin{minipage}{9.5em}
          \resizebox{\textwidth}{!}{
            \begin{tikzpicture}[scale=1.8]
              % Graph nodes
              \tikzstyle{VertexStyle} = [shape = circle, minimum width
              = 3em, fill=lightgray]
              \Vertex[Math,L=Y,x=0,y=0]{Y}
              \Vertex[Math,L=X^1,x=-1.5,y=0.5]{X1}
              \Vertex[Math,L=X^2,x=-1.5,y=-0.5]{X2}
              \tikzstyle{VertexStyle} = [shape = circle, minimum
              width=3em, color=black, fill=col1]
              \Vertex[Math,L=I,x=-3,y=0]{I2}
              % Graph edges
              \tikzstyle{EdgeStyle} = [-Latex, line width=1]
              \Edge(X1)(Y)
              \Edge(Y)(X2)
              \tikzset{EdgeStyle/.append style = {color=black}}
              \Edge(I2)(X2)
            \end{tikzpicture}}
        \end{minipage}
      };
      \node[rectangle, draw] (data2) at (-2.4, -1.2) {%
        \begin{minipage}{9.5em}
          \resizebox{\textwidth}{!}{
            \begin{tikzpicture}[scale=1.8]
              % Graph nodes
              \tikzstyle{VertexStyle} = [shape = circle, minimum width
              = 3em, fill=lightgray]
              \Vertex[Math,L=Y,x=0,y=0]{Y}
              \Vertex[Math,L=X^1,x=-1.5,y=0.5]{X1}
              \Vertex[Math,L=X^2,x=-1.5,y=-0.5]{X2}
              \tikzstyle{VertexStyle} = [shape = circle, minimum
              width=3em, color=black, fill=col2]
              \Vertex[Math,L=I,x=-3,y=0]{I2}
              % Graph edges
              \tikzstyle{EdgeStyle} = [-Latex, line width=1]
              \Edge(X1)(Y)
              \Edge(Y)(X2)
              \tikzset{EdgeStyle/.append style = {color=black}}
              \Edge(I2)(X2)
            \end{tikzpicture}}
        \end{minipage}
      };
      \node[rectangle, draw] (data3) at (4.75, -1.2) {%
        \begin{minipage}{9.5em}
          \resizebox{\textwidth}{!}{
            \begin{tikzpicture}[scale=1.8]
              % Graph nodes
              \tikzstyle{VertexStyle} = [shape = circle, minimum width
              = 3em, fill=lightgray]
              \Vertex[Math,L=Y,x=0,y=0]{Y}
              \Vertex[Math,L=X^1,x=-1.5,y=0.5]{X1}
              \Vertex[Math,L=X^2,x=-1.5,y=-0.5]{X2}
              \tikzstyle{VertexStyle} = [shape = circle, minimum
              width=3em, color=black, fill=col3]
              \Vertex[Math,L=I,x=-3,y=0]{I2}
              % Graph edges
              \tikzstyle{EdgeStyle} = [-Latex, line width=1]
              \Edge(X1)(Y)
              \Edge(Y)(X2)
              \tikzset{EdgeStyle/.append style = {color=black}}
              \Edge(I2)(X2)
            \end{tikzpicture}}
        \end{minipage}
      };
    \end{tikzpicture}}
  \caption{Illustration of multi-environment data generation
    setting. Only some environments are observed, but one would like
    to be able to make predictions on any further potentially
    unobserved environment.}\label{fig:env_conditions}
\end{figure}

An illustration of the multi-environment regression setting is given
in Figure~\ref{fig:env_conditions}.
Neglecting the environment structure, a classical approach to this
problem is to use least squares to estimate a function
$f:\bm{\mathcal{X}}\rightarrow\R$ which minimizes the (weighted)
pooled squared loss
\begin{equation}
  \label{eq:regression_pooled}
  \frac{1}{n}\sum_{e\in\mathcal{E}^{\text{obs}}}n_e\cdot\E\left((Y_e-f(X_e))^2\right),
\end{equation}
where $n_e$ is the number of observations in environment $e$. Due to the
heterogeneity, the optimizer on each individual environment, which is
given by $f_e(x)=\E(Y_e\,\vert\, X_e=x)$, generally changes across
environments. Therefore, it is not necessarily the case that the
pooled optimizer generalizes to unseen settings
$e\in\mathcal{E}\setminus\mathcal{E}^{\text{obs}}$. Instead, we
propose to explicitly use the assumed invariance in
\eqref{eq:invariant_mechanism_assumption} and estimate a function
$f:\bm{\mathcal{X}}\rightarrow\R$ which minimizes the pooled squared
loss in \eqref{eq:regression_pooled} subject to the constraint that
there exists a subset $S\subseteq\{1,\dots,d\}$ such that for all
$e\in\mathcal{E}$ and all $x\in\bm{\mathcal{X}}$ it holds that
\begin{equation}
  \label{eq:stability_constraint}
  f(x)=\E(Y_e\,\vert\, X_e^{S}=x^S).
\end{equation}
Define the constraint set
$\mathcal{C}=\{f:\bm{\mathcal{X}}\rightarrow\R\,\vert\, f \text{
  satisfies \eqref{eq:stability_constraint}}\}$, which is non-empty by
the assumption in
\eqref{eq:invariant_mechanism_assumption}. Therefore, we have the
following well-defined optimization problem
\begin{equation}
  \label{eq:constrained_opt}
  \text{minimize}\quad\frac{1}{n}\sum_{e\in\mathcal{E}^{\text{obs}}}n_e\cdot\E\left((Y_e-f(X_e))^2\right)\quad\text{subject
    to } f\in\mathcal{C}.
\end{equation}
The standard approach to this problem is to solve this optimization
directly by optimizing over the function $f$. However, we suggest a
different approach. The optimization problem in
\eqref{eq:constrained_opt} is equivalent to searching over all subset
$S\subseteq\{1,\dots,d\}$ which satisfy
\eqref{eq:stability_constraint} and for which the conditional mean
based on the predictors $X^S$ has minimal loss in
\eqref{eq:regression_pooled}. The solution to the optimization is then
simply the conditional mean based on $X^S$. Such a set is not
necessarily unique which is why our proposed method in
Section~\ref{sec:meth_details} averages over an estimate of all these
sets. The reason we use this optimization approach is that, in
particular in the finite sample case, the averaging technique leads to
improved performance. This can be seen in
Sections~\ref{sec:numerical_sim}~and~\ref{sec:application_bio} in
comparison with the instrumental variable procedure that in the linear
case directly optimizes \eqref{eq:constrained_opt}. In
Example~\ref{ex:toy_example}, we illustrate the difference between the
unconstrained optimization in \eqref{eq:regression_pooled} and
constrained optimization in \eqref{eq:constrained_opt} based on a toy
example.
\begin{example}[toy model]\label{ex:toy_example}
  Consider a variable $I$ which generates the environments or
  perturbations. Let the variables $(I, X, Y)$ satisfy the following
  structural
  causal model (Definition~\ref{def:SCM})\\
  \begin{minipage}{0.5\textwidth}
    \begin{equation*}
      \mathcal{S}^*\quad
      \begin{cases}
        I\phantom{I^1}\coloneqq \varepsilon_{I}\\
        X^1\coloneqq \varepsilon_{X^1}\\
        Y\phantom{^1}\coloneqq X^1 + \varepsilon_{Y}\\
        X^2\coloneqq Y + I+\varepsilon_{X^2}\\
        X^3\coloneqq Y + I+\varepsilon_{X^3}\\
      \end{cases}
    \end{equation*}
   \end{minipage}%
  \begin{minipage}{0.5\textwidth}
    \resizebox{0.8\textwidth}{!}{
      \begin{tikzpicture}[scale=1.6]
        % Graph nodes
        \tikzstyle{VertexStyle} = [shape = circle, minimum width =
        3em, fill=lightgray]
        \Vertex[Math,L=Y,x=0,y=0]{Y}
        \Vertex[Math,L=X^1,x=-1.5,y=0.6]{X1}
        \Vertex[Math,L=X^2,x=-1.5,y=-0.6]{X2}
        \Vertex[Math,L=X^3,x=1.5,y=-0.6]{X3}
        \tikzstyle{VertexStyle} = [draw, shape = circle, minimum
        width=3em, text=red, color=red, fill=lightgray]
        \Vertex[Math,L=I,x=-3,y=0]{I2}
        % Graph edges
        \tikzset{EdgeStyle/.append style = {-Latex, line width=1}}
        \Edge(X1)(Y)
        \Edge(Y)(X2)
        \Edge(Y)(X3)
        \tikzset{EdgeStyle/.append style = {color=red}}
        \Edge(I2)(X2)
      \end{tikzpicture}}
  \end{minipage}
  
  \noindent with $\varepsilon_{Y}$,
  $\varepsilon_{X^1},\varepsilon_{X^2}$ and $\varepsilon_{X^3}$
  independently $\mathcal{N}(0,1)$-distributed and
  $\varepsilon_{I}=c(e)$ for a constant $c(e)\in\R$ depending on the
  environment $e\in\mathcal{E}^{\text{obs}}$. Variable $I$ is
  unobserved and describes the changes across environments (see
  Section~\ref{sec:SP_vars}). Consider two cases, where (i) only the
  variables $(Y, X^1, X^2)$ and (ii) only the variables
  $(Y, X^1, X^2, X^3)$ are observed.  Given case (i) and assuming a
  mixture model across the observed environments
  $\mathcal{E}^{\text{obs}}$ (with equal probabilities across all
  environments) allows us to compare optimization of
  \eqref{eq:regression_pooled} solved by a pooled least squares
  estimator with optimization \eqref{eq:constrained_opt} by a simple calculation. The standard
  ordinary least squares (OLS) estimator in the population case is
  given by
  \begin{equation*}
    \beta^{\operatorname{OLS}}
    =
    \begin{pmatrix}
      \var(X^1) &\cov(X^1, X^2)\\
      \cov(X^1, X^2) &\var(X^2)
    \end{pmatrix}^{-1}
    \begin{pmatrix}
      \cov(X^1, Y)\\
      \cov(X^2, Y)
    \end{pmatrix}
    =
    \begin{pmatrix}
      \frac{1+\var(I)}{2+\var(I)}\\
      \frac{1}{2+\var(I)}
    \end{pmatrix},
  \end{equation*}
  where by slight abuse of notation $\var(I)$ refers to the variation
  of $c(e)$ across environments. Hence, the coefficient of $X^2$ is
  non-zero in this case implying that predictions can become bad on
  environments where $I$ takes large values. Since the constraint in
  \eqref{eq:stability_constraint} is satisfied for both
  $S=\varnothing$ and $S=\{1\}$, the optimizer of
  \eqref{eq:constrained_opt} is given by
  $f(x)=\E(Y\,\vert\, X^1=x^1)=x^1$ and the optimal regression
  parameter is given by $\beta^*=(1, 0)^{\top}$. This regression
  coefficient is ideal in the sense that it contains all the
  information about $Y$ that can be explained independent of the value
  of $I$. If the observed perturbations have a large spread, i.e.,
  $\frac{1}{\abs{\mathcal{E}^{\text{obs}}}}\sum_{e\in\mathcal{E}^{\text{obs}}}c(e)^2$
  is large, then the OLS regression parameter
  $\beta^{\operatorname{OLS}}$ approximates the constrained regression
  parameter $\beta^{*}$ (see Corollary~\ref{thm:optimal_OLS}). Strong
  heterogeneity in the data therefore improves the generalization
  performance of a standard pooled regression.

  Consider now case (ii), in which we additionally observe variable
  $X^3$. While $X^2$ was harmful for the generalization performance,
  $X^3$ is in general beneficial (see
  Figure~\ref{fig:illustrative}). In particular, the regression
  parameter for the regression of $Y$ on $(X^1,X^2,X^3)$ with the
  constraint in \eqref{eq:stability_constraint} has the form
  $\beta^{*}=(\beta^*_1,0,\beta^*_2)$, where the two parameters are in
  general non-zero and depend on the underlying system. Similar to
  case (i), it can be shown that the standard OLS parameter again
  converges to this constrained estimator if the interventions are
  sufficiently strong. A formal result describing when the pooled OLS
  converges to the constrained optimizer in the case of linear systems
  is given in Section~\ref{sec:OLS_behavior}. In many application,
  however, there might be insufficient heterogeneity for the OLS and
  the difference between solutions to \eqref{eq:regression_pooled} and
  \eqref{eq:constrained_opt} might be substantial. Therefore, whenever
  the training environments consist of weaker interventions than the
  testing environment, one can benefit from explicitly incorporating
  stability into the estimation (also shown in
  Figure~\ref{fig:illustrative}).
\end{example}
The pooled squared loss \eqref{eq:regression_pooled} and the constraint
\eqref{eq:stability_constraint} combine two aspects: (i)
Predictive performance of the model given by the optimization
objective and (ii) stability across perturbations enforced by the
constraint in \eqref{eq:stability_constraint}. These concepts are
formalized in the following definitions.
\begin{definition}[generalizable sets]
  \label{def:generalizable}
  A set $S\subseteq\{1,\dots, d\}$ is called generalizable with
  respect to $\widetilde{\mathcal{E}}\subseteq\mathcal{E}$ if for all
  $e,f\in\widetilde{\mathcal{E}}$ and for all $x\in\bm{\mathcal{X}}$
  it holds that
  \begin{equation*}
    \E\left(Y_e\vert X^S_e =x^S\right)=\E\left(Y_f\vert X^S_f=x^S \right).
  \end{equation*}
  We denote by $\mathbb{G}_{\widetilde{\mathcal{E}}}$ the collection
  of all generalizable sets.
\end{definition}
Any generalizable set will by definition have the property that a
regression based on the predictors in that set should have similar
predictive performance across all environments $e\in\widetilde{\mathcal{E}}$. In
practice, it is, however, also important that the predictive
performance is not only equal across different environments but is
equally good in all environments.
\begin{definition}[generalizable and regression optimal sets]
  \label{def:regopt}
  A set $S\subseteq\{1,\dots,d\}$ is called generalizable and
  regression optimal with respect to
  $\widetilde{\mathcal{E}}\subseteq\mathcal{E}$ if it is generalizable
  in the sense that $S\in\mathbb{G}_{\widetilde{\mathcal{E}}}$ and if it satisfies
  \begin{equation*}
    S\in\argmin_{\bar{S}\in\mathbb{G}_{\widetilde{\mathcal{E}}}}\E[(Y-\E(Y \vert
    X^{\bar{S}}))^2].
  \end{equation*}
  Here, the expectation is with respect to the mixture distribution
  over all distributions in $\widetilde{\mathcal{E}}$.  The collection
  of all generalizable and regression optimal sets is denoted by
  $\mathbb{O}_{\widetilde{\mathcal{E}}}$.
\end{definition}
In general, the sizes of the collections
$\mathbb{G}_{\widetilde{\mathcal{E}}}$ and
$\mathbb{O}_{\widetilde{\mathcal{E}}}$ decrease when more environments
are added to $\widetilde{\mathcal{E}}$. In
Section~\ref{sec:identifiability}, we discuss when the observed
environments $\mathcal{E}^{\text{obs}}$ are sufficient for
generalization on all potential environments $\mathcal{E}$, i.e., when
$\mathbb{G}_{\mathcal{E}^{\text{obs}}}=\mathbb{G}_{\mathcal{E}}$ and
$\mathbb{O}_{\mathcal{E}^{\text{obs}}}=\mathbb{O}_{\mathcal{E}}$
hold.

Generalizable and regression optimal sets are the main focus of our
paper. In Section~\ref{sec:meth_details}, we will introduce an
algorithm that approximates a solution to the constrained optimization
\eqref{eq:constrained_opt}, by explicitly estimating the generalizable
and regression optimal sets.

\section{Stable blankets}\label{sec:SP_vars}

Previously, we did not assume an underlying causal model. This is
sufficient whenever we are only interested in stable prediction across
environments. Here, we make additional assumptions on the underlying
model, which allow us to specify graphical conditions for computing
generalizable sets. This characterization is not important from a
methodological viewpoint but helps from a causal modeling perspective
and can give some useful insights that help interpret the results of
variable selection. It uses some terminology and concepts from the
causal literature. The practical oriented reader might skip this
subsection.

We choose to work with structural causal models (SCMs)
\citep[e.g.,][]{Pearl2009,Jonasbook}, sometimes also referred to as
structural equation models (SEMs).
\begin{definition}[structural causal model]
  \label{def:SCM}
  A structural causal model (SCM), over random variables
  $W=(W^1, \ldots W^p)$ is a collection of $p$ assignments
  \begin{equation}
    \label{eq:scmiid}
    \mathcal{S}\quad
    \begin{cases}
      W^1\coloneqq f^1(W^{\PA(W^1)}, \varepsilon^1)\\
      \qquad \vdots\\
      W^p\coloneqq f^d(W^{\PA(W^d)}, \varepsilon^p),
    \end{cases}
  \end{equation}
  where $\varepsilon^1, \ldots, \varepsilon^p$ are independent noise
  variables. For all $k \in \{1, \ldots, p\}$,
  $\PA(W^k) \subseteq \{1, \ldots, p\}\setminus\{k\}$ is called the
  set of direct (causal) parents of $W^k$.  Moreover, the assignments
  in \eqref{eq:scmiid} are assumed to be uniquely solvable, which is
  always true if the induced graph is acyclic, for example. An SCM
  induces a distribution over the variables $W$ as well as a graph
  over the vertices $(W^1, \ldots, W^p)$, denoted by
  $\mathcal{G}(\mathcal{S})$, by adding directed edges from $\PA(W^k)$
  to $W^k$ for all $k \in \{1, \ldots, p\}$.
\end{definition}
For any SCM $\mathcal{S}$ over $W=(W^1,\dots, W^p)$, an intervention
on a variable $W^j$ corresponds to a new SCM $\tilde{S}$ for which
only the structural assignment of $W^j$ has been replaced. Note that
we only consider interventions for which the new SCM remains
solvable. When talking about graphs we use the notion of d-separation
\citep[e.g.,][]{Pearl2009}, which we denote by $\independentG$ to
distinguish it from conditional independence. We summarize the causal
model setting below.
\begin{setting}[underlying causal model]
  \label{setting:causal}
  Let
  $X\in\bm{\mathcal{X}}=\mathcal{X}^1\times\dots\times\mathcal{X}^d$
  be predictor variables, $Y\in\R$ a response variable and
  $I=(I^1,\ldots,I^m)\in\bm{\mathcal{I}}=\mathcal{I}^1\times\dots\times\mathcal{I}^d$
  intervention variables which are assumed to be unobserved and are
  used to formalize interventions. Assume there exists a fixed SCM
  $\mathcal{S}^*$ over $(I, X, Y)$ such that
  $\mathcal{G}(\mathcal{S}^*)$ is a directed acyclic graph (DAG) and
  for which the intervention variables $I$ are source nodes and do not
  appear in the structural assignment of $Y$. Let $\mathcal{E}$
  consist of all intervention environments, where for every
  $e\in\mathcal{E}$ there is an intervention SCM $\mathcal{S}_e$ over
  $(I_e,X_e,Y_e)$ in which only equations with $I_e$ on the right-hand
  side change and the graph structure stays fixed (i.e.,
  $\mathcal{G}(\mathcal{S}_e)=\mathcal{G}(\mathcal{S}^*)$ for all
  $e\in\mathcal{E}$). Assume the distribution of $(I_e, X_e, Y_e)$ is
  absolutely continuous with respect to a product measure that
  factorizes. Lastly, let
  $\mathcal{E}^{\text{obs}}\subseteq\mathcal{E}$ be a finite set of
  observed environments.
\end{setting}
Based on this setting we can define intervention stable
sets. Intuitively, a set $S$ is called intervention stable if the
corresponding predictors explain all of the intervention variability
in the response variable.
\begin{definition}[intervention stable sets]
  Given Setting~\ref{setting:causal}, a set $S\subseteq\{1,\dots, d\}$
  is called intervention stable if for all $\ell\in\{1,\dots,m\}$ the
  d-separation $I^{\ell} \independentG Y \,\vert\, X^S$ holds in
  $\mathcal{G}(S^{*})$.
\end{definition}
Since the graph remains fixed across interventions, it immediately
follows that the parents of the response $\PA(Y)$ are an intervention
stable set. Together with the following proposition (which proves
that any intervention stable set is generalizable) this implies that
the invariance assumption in \eqref{eq:invariant_mechanism_assumption}
is satisfied.
\begin{proposition}[intervention stable sets are generalizable]
  \label{thm:invariance_equiv}
  Assume Setting~\ref{setting:causal}, then for all intervention
  stable sets $S\subseteq\{1,\dots, d\}$ it holds that
  $S\in\mathbb{G}_{\mathcal{E}}$.
\end{proposition}
A proof is given in Appendix~\ref{sec:proofs}. Based on this
proposition it is possible to find generalizable sets using only the
graphical structure. However, not all generalizable sets are
intervention stable. More details on this relation are given in
Section~\ref{sec:identifiability}.

In graphical models the Markov blanket of $Y$, denoted by $\MB(Y)$, is
defined as the smallest set $S\subseteq\{1,\dots,d\}$ that satisfies
\begin{equation*}
  \forall j\in\{1,\dots,d\}\setminus S:\quad X^j\independentG Y \,\vert\, X^{S}.
\end{equation*}
The Markov blanket specifies the smallest set of variables that
separates the response $Y$ from all other variables and hence allows a
precise notion of predictiveness. The following definition combines
this notion with intervention stability.\footnote{The union of
  intervention stable sets is itself not necessarily intervention
  stable anymore.}
\begin{definition}[stable blanket]
  \label{def:stable_blanket}
  Assume Setting~\ref{setting:causal} and define the following set of variables
  \begin{equation*}
    N^{\text{int}}\coloneqq\{1,\dots,d\}\setminus\{j\in\{1,\dots,d\}\,\vert\,
    \exists k\in\CH^{\operatorname{int}}(Y): j\in\DE(X^k)\},
  \end{equation*}
  where $\CH^{\operatorname{int}}(Y)$ are all children of $Y$ that are
  directly intervened on and $\DE(X^k)$ are all descendants of $X^k$
  including $X^k$ itself. Then, the stable blanket, denoted by
  $\SB(Y)$, is defined as the smallest set $S\subseteq N^{\text{int}}$
  that satisfies
  \begin{equation*}
    \forall j\in N^{\text{int}}\setminus S:\quad X^j\independentG Y \,\vert\, X^{S}.
  \end{equation*}
\end{definition}
A different characterization of the stable blanket is given in the
following theorem, which also proves that the stable blanket is
generalizable and regression optimal.
\begin{theorem}[stable blankets are generalizable and regression
  optimal]
  \label{thm:stable_blanket}
  Assume Setting~\ref{setting:causal}, then the stable blanket
  consists of all children of $Y$ that are in $N^{\text{int}}$,
  the parents of such children and the parents of $Y$. Furthermore, it
  holds that $\SB(Y)\in\mathbb{O}_{\mathcal{E}}$.
\end{theorem}
A proof is given in Appendix~\ref{sec:proofs}. It is illustrative to
think about the set $\SB(Y)$ in relation to the parent set $\PA(Y)$
and the Markov blanket $\MB(Y)$. By
Theorem~\ref{thm:stable_blanket}, it will lie somewhere between
these two sets. The exact size depends on the intervention variables,
with the following special cases: (i) if there are no interventions it
holds that $\SB(Y)=\MB(Y)$, (ii) if there are sufficiently many
interventions, e.g., on any node other than $Y$, it holds that

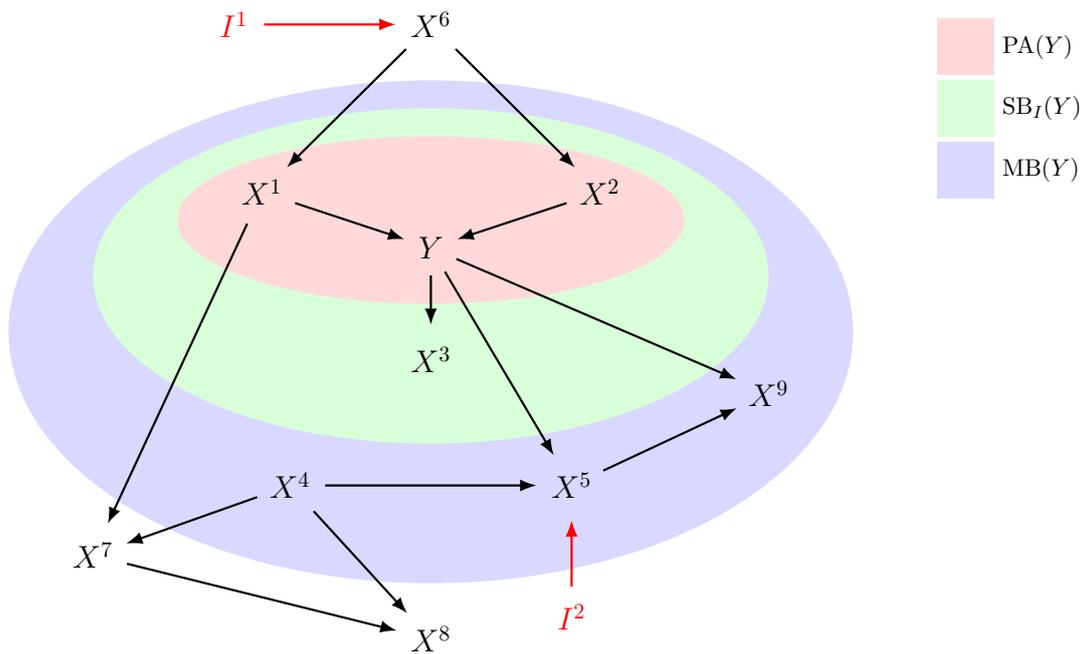
\begin{figure}[ht]
  \centering
    \resizebox{0.9\textwidth}{!}{
      \begin{tikzpicture}[scale=0.9]
        \def\firstellipse{(0,2) ellipse (4.5 and 1.5)}
        \def\secondellipse{(0,1) ellipse (6 and 3)}
        \def\thirdellipse{(0,0) ellipse (7.5 and 4.5)}
        \def\fourthellipse{(0,0) ellipse (10 and 7)}
        \fill[fill=blue!15]\thirdellipse;
        \fill[fill=green!15]\secondellipse;
        \fill[fill=red!15]\firstellipse;
        % Graph nodes
        \SetGraphUnit{2}
        \tikzstyle{VertexStyle} = [shape = circle, minimum width = 2em]
        \Vertex[Math,L=Y,x=0,y=1.5]{Y}
        \Vertex[Math,L=X^1,x=-3,y=2.5]{X1}
        \Vertex[Math,L=X^2,x=3,y=2.5]{X2}
        \Vertex[Math,L=X^3,x=0,y=-0.5]{X3}
        \Vertex[Math,L=X^4,x=-2.5,y=-2.5]{X4}
        \Vertex[Math,L=X^5,x=2.5,y=-2.5]{X5}
        \Vertex[Math,L=X^6,x=0,y=5.5]{X6}
        \Vertex[Math,L=X^7,x=-6,y=-4]{X7}
        \Vertex[Math,L=X^8,x=0,y=-5.5]{X8}
        \tikzstyle{VertexStyle} = [shape = circle, minimum width =
        2em, text=red]
        \Vertex[Math,L=I^1,x=-3.5,y=5.5]{I1}
        \Vertex[Math,L=I^2,x=2.5,y=-5.1]{I2}
        % Graph edges
        \tikzset{EdgeStyle/.append style = {-Latex, line width=1}}
        \Edge(X1)(Y)
        \Edge(X2)(Y)
        \Edge(Y)(X3)
        \Edge(Y)(X5)
        \Edge(X4)(X5)
        \Edge(X6)(X1)
        \Edge(X6)(X2)
        \Edge(X1)(X7)
        \Edge(X7)(X8)
        \Edge(X4)(X8)
        \Edge(X4)(X7)
        \tikzset{EdgeStyle/.append style = {color=red}}
        \Edge(I1)(X6)
        \Edge(I2)(X5)
        % Set names
        \draw[fill=red!15, draw=none] (9,4.6) rectangle (10,5.6);
        \draw[fill=green!15, draw=none] (9,3.5) rectangle (10,4.5);
        \draw[fill=blue!15, draw=none] (9,2.4) rectangle (10,3.4);
        \node[anchor=west] at (10, 5.1) {$\PA(Y)$};
        \node[anchor=west] at (10, 4) {$\SB(Y)$};
        \node[anchor=west] at (10, 2.9) {$\MB(Y)$};
      \end{tikzpicture}}
    \caption{Graphical illustration of variable selection. The goal is
      to find predictors $X=(X^1,\dots,X^{9})$ that are functionally
      related to the response $Y$. Here, variables $I=(I^1,I^2)$ are
      unobserved intervention variables. The colored areas represent
      different targets of inference: Markov blanket, stable blanket
      and parents (causal variables). If the goal is to get as close
      as possible to the parents, the stable blanket can improve on
      the Markov blanket if there are sufficiently many informative
      interventions.}
  \label{fig:stablepred}
\end{figure}

$\SB(Y)=\PA(Y)$. A visualization of these relations is given in
Figure~\ref{fig:stablepred}. Whenever the goal is to find the direct
causal parents of a response variable, one can generally get closer than
the Markov blanket by considering the stable blanket.

\subsection{Stable blanket as a proxy for
  causality}\label{sec:variable_selection}

As alluded to in the previous section, the stable blanket $\SB(Y)$ can
be related to an underlying causal model. In the most basic case of an
SCM with an underlying directed acyclic structure, the Markov blanket
can be decomposed into parents, children and parents of children, i.e.,
\begin{equation}
  \label{eq:Markov_decomp}
  \MB(Y)=\PA(Y) \cup \CH(Y) \cup \{j\in\{1,\dots,d\}\,\vert\, \exists
  k\in\CH(Y): j\in\PA(X^k)\}.
\end{equation}
As long as the intervention variables do not directly affect the
response $Y$ this implies that the difference between the Markov blanket
and the stable blanket consists only of variables that are children or
parents of children of the response. We denote this difference as the non-stable
blanket
\begin{equation*}
  \NSB(Y)\coloneqq\MB(Y)\setminus\SB(Y).
\end{equation*}
It can be shown that any set containing variables from the non-stable
blanket are not intervention stable. Given the decomposition in
\eqref{eq:Markov_decomp}, this implies that $\PA(Y)\subseteq\SB(Y)$
and
$\NSB(Y)\subseteq\CH(Y) \cup \{j\in\{1,\dots,d\}\,\vert\, \exists
k\in\CH(Y): j\in\PA(X^k)\}$.  Therefore, depending on whether we are
either interested in the parents or in down-stream variables (or
children) of $Y$, the sets $\SB(Y)$ and $\NSB(Y)$ can be used as
proxies.

\subsection{Identifiability of generalizable
  sets}\label{sec:identifiability}

In Section~\ref{sec:generalizing_regression}, we introduced the
collection of generalizable and regression optimal predictor sets
$\mathbb{O}_{\mathcal{E}}$ which lead to regressions that behave well
on all potential environments $\mathcal{E}$. We saw that if one
assumes an underlying causal model, as in Section~\ref{sec:SP_vars},
it is possible to compute the stable blanket $\SB(Y)$. This shows,
since $\SB(Y)\in\mathbb{O}_{\mathcal{E}}$, that it is possible to
construct a generalizable and regression optimal set whenever the
underlying causal structure is known. In practice, we usually do not
have access to the causal structure and only observe a (small) subset
$\mathcal{E}^{\text{obs}}$ of all potential environments
$\mathcal{E}$. Intuitively, the best one can hope for in such cases is
to find sets in $\mathbb{O}_{\mathcal{E}^{\text{obs}}}$. Therefore,
the question arises whether and when the sets in
$\mathbb{O}_{\mathcal{E}^{\text{obs}}}$ also generalize to any further
environments not contained in $\mathcal{E}^{\text{obs}}$. The answer
depends on the assumptions one is willing to make on the data
generating process and, in particular, on the types of environments that
are observed and unobserved. In this section, we discuss
additional conditions to Setting~\ref{setting:causal}, that allow
generalization from $\mathcal{E}^{\text{obs}}$ to $\mathcal{E}$.

Given Setting~\ref{setting:causal}, we are interested in what
additional conditions are sufficient to be able to infer the stable
blanket and hence a generalizable and regression optimal set from
data. As $\SB(Y)$ is defined as the union of intervention stable sets
we need to be able to determine whether a given set satisfies this
property based on data. We require two types of assumptions.

Firstly, the faithfulness assumption \citep{Pearl2009} ensures that
any d-separation in the graph corresponds to a conditional
independence in the data generating random variables. Given
faithfulness and a sufficiently large sample size it is possible in
most cases to consistently recover the Markov blanket using, for
example, an appropriate feature selection algorithm
\citep{pellet2008using}. This, in particular, does not require any
type of heterogeneity and can be based purely on observational
data.

Secondly, to check whether a subset $S\subseteq\{1,\dots,d\}$ is
intervention stable requires to detect all conditional dependencies
between the intervention variables and the response given the
predictors in $S$. Since only the environments are observed and not
the intervention variables, we require that
\begin{equation*}
  \forall e,f\in\mathcal{E}^{\text{obs}}:
  \E(Y_e\,\vert\, X_e^{S}=x^S)=\E(Y_f\,\vert\, X_f^{S}=x^S)
  \quad\Rightarrow\quad
  \forall\ell\in\{1,\dots,m\}: I^{\ell} \independent Y \vert X^S.
\end{equation*}
In other words, by contraposition, we need that any conditional dependence between the
intervention variables and the response leads to a shift in conditional
mean across environments, i.e., we need to observe sufficiently many
informative environments.

\subsection{Understanding stable blankets in linear models}\label{sec:OLS_behavior}

To get a better understand of the relation between stable blankets and
standard regression techniques, we consider linear models and analyze
the behavior of the pooled ordinary least squares (OLS) estimator in
our proposed multi-environment regression setting. We will show that
in order for the OLS only sets variables in the non-stable blanket to
zero if the intervention strength goes to infinity. This means that
whenever the intervention strength is not sufficiently strong, OLS
does not necessarily perform well on unobserved environments with
stronger interventions.

For our results, it is enough to consider population quantities since
the ordinary least squares estimator is consistent. The following
lemma gives an explicit expression of the population OLS applied to a
linear SCM in terms of the (exogenous) noise variables and the
structure matrix. It allows us to assess the behavior of the OLS under
interventions.
\begin{lemma}[OLS in linear SCMs]
  \label{thm:OLS_SCM}
  Assume the variables $(X, Y)\in\R^{d+1}$ satisfy a linear directed
  acyclic SCM, i.e., there exists $B\in\R^{(d+1)\times (d+1)}$ and
  independent noise variable
  $\epsilon=(\epsilon^{0},\dots \epsilon^{d})\in\R^{d+1}$ such that
  \begin{equation*}
    \begin{pmatrix}
      Y\\X
    \end{pmatrix}
    \coloneqq B\cdot
    \begin{pmatrix}
      Y\\X
    \end{pmatrix}
    + \epsilon
    \quad\text{with}\quad
    B=
    \begin{pmatrix}
      0 &\beta_{\tPA}^{\top}\\
      \beta_{\tCH} &B_X
    \end{pmatrix}.
  \end{equation*}
  The parents and children of $Y$ are given by the non-zero coefficients
  of $\beta_{\tPA}$ and $\beta_{\tCH}$, respectively. Then the
  population ordinary least squares $\beta^{\operatorname{OLS}}$, when
  regressing $Y$ on $X$, is given by
  \begin{equation*}
    \beta^{\operatorname{OLS}}=\beta_{\tPA}+
      \left((\vI-B_{X})^{\top} - \beta_{\tPA}\beta_{\tCH}^{\top}\right)D^{-1}\beta_{\tCH}\left(1-
      \frac{\var(\epsilon^{0})\beta_{\tCH}^{T}D^{-1}\beta_{\tCH}}{1+\var(\epsilon^0)\beta_{\tCH}^{T}D^{-1}\beta_{\tCH}}\right)\var(\epsilon^0),
  \end{equation*}
  where $D=\cov(\epsilon^1,\dots,\epsilon^d)$.
\end{lemma}
A proof is given in Appendix~\ref{sec:proofs}. The result implies that
the population OLS can be decomposed into the sum of the true causal
parameter $\beta_{\tPA}$ plus a correction term. It can be shown that
this correction is zero for coordinates $j\not\in\MB(Y)$ (see proof of
Corollary~\ref{thm:optimal_OLS}), which is a well-known property of
ordinary least squares. The result allows to explicitly analyze the
behavior of the OLS in the multi-environment regression setting. In
particular, it can be shown that $\beta^{\operatorname{OLS},j}$
converges to zero for variables $j\not\in\SB(Y)$ as the variance of
the interventions across environments increases. The exact result is
given in the following corollary.

\begin{corollary}[OLS under strong interventions]
  \label{thm:optimal_OLS}
  Let $(I_{\bar{n}}, X_{\bar{n}}, Y_{\bar{n}})$ be a sequence of variables satisfying
  Setting~\ref{setting:causal} for the same directed acyclic linear
  SCM $\mathcal{S}^*$. Additionally, assume that each of the variables
  $I_n$ has exactly one child and the sum of the coefficients along
  directed paths starting at variables $I_{\bar{n}}$ are always
  non-vanishing. Moreover, for all $\bar{n}\in\N$ there are two observed
  environments $\mathcal{E}^{\text{obs}}_{\bar{n}}=\{e^{+}_{\bar{n}},e^{-}_{\bar{n}}\}$, where
  the interventions $e^{+}_{\bar{n}}$ and $e^{-}_{\bar{n}}$ satisfy for all
  $\ell\in\{1,\dots,m\}$ that
  \begin{equation*}
    I^{\ell}_{\bar{n}}=c_{\ell,\bar{n}}^{+}\text{ in }\mathcal{S}_{e^{+}_{\bar{n}}}
    \quad\text{and}\quad
    I^{\ell}_{\bar{n}}=c_{\ell,\bar{n}}^{-}\text{ in }\mathcal{S}_{e^{-}_{\bar{n}}},
  \end{equation*}
  where $c_{\ell,\bar{n}}^{+},c_{\ell,\bar{n}}^{-}$ are independent
  random variables with mean zero and variance $\sigma_{\bar{n}}^2$
  such that
  $\lim_{\bar{n}\rightarrow\infty}\sigma_{\bar{n}}=\infty$. Then, the
  pooled OLS estimator $\beta^{\operatorname{OLS}}_{\bar{n}}$ when
  regressing $Y_{\bar{n}}$ on $X_{\bar{n}}$ (i.e., the minimizer of
  \eqref{eq:regression_pooled} over all linear functions) satisfies
  for all $j\in\{1,\dots,d\}\setminus\SB(Y)$,
  \begin{equation*}
    \lim_{\bar{n}\rightarrow\infty}\beta^{\operatorname{OLS},j}_{\bar{n}}=0.
  \end{equation*}
\end{corollary}
A proof is given in Appendix~\ref{sec:proofs}. We use $\bar{n}$ to
make clear that this is a population result in which the limit is
taken in terms of intervention strength and not in terms of sample
size. Corollary~\ref{thm:optimal_OLS} provides results in an
asymptotic regime in which the interventions are sufficiently strong.
In the numerical simulations in Section~\ref{sec:numerical_sim}, we
will see that whenever the intervention strength is not sufficiently
strong, the OLS can be outperformed.

\section{Proposed method}\label{sec:meth_details}

Our goal is to fit a regression function, which approximates a
solution to \eqref{eq:constrained_opt}. Instead of just finding a
single set $S$ for which the conditional mean based on $X^S$ solves
\eqref{eq:constrained_opt}, we propose to approximate this function
with a weighted average. The idea is that verifying the invariance
constraint in \eqref{eq:stability_constraint} involves uncertainty
which can be reduced by averaging over many invariant sets instead of
deciding on a single set. For any subset $S\subseteq\{1,\dots, d\}$,
let $\hat{f}^{S}:\bm{\mathcal{X}}^{\abs{S}}\rightarrow\R$ be a
regression estimate which minimizes \eqref{eq:regression_pooled}
restricted to the predictors in $S$. Recall that, the \emph{stabilized
  regression} estimator is defined as the weighted average
\begin{equation}
  \label{eq:stabilizedregression_estimator2}
  \hat{f}_{\operatorname{SR}}(X)\coloneqq\sum_{S\subseteq\{1,\dots,
    d\}}\widehat{w}_S\cdot \hat{f}^{S}(X^{S}),
\end{equation}
where the weights are assumed to satisfy $\sum_S \widehat{w}_S=1$. For
this estimator to approximate a solution of
\eqref{eq:constrained_opt}, we select large weights for sets of
predictors which are both generalizable and regression optimal.

\subsection{Estimating generalizable and regression optimal
  sets}\label{sec:weights}

Let $\widehat{\mathbb{O}}\subseteq\mathcal{P}(\{1,\dots,d\})$ of the
collection of generalizable and regression optimal sets with respect
to $\mathcal{E}^{\text{obs}}$. Then, we propose to construct the
weights as follows,
\begin{equation}
  \label{eq:weights}
  \widehat{w}_S\coloneqq
  \begin{cases}
    1/\abs{\widehat{\mathbb{O}}}\quad &\text{if}\quad S\in\widehat{\mathbb{O}}\\
    0\quad &\text{otherwise.}
  \end{cases}
\end{equation}
The set $\widehat{\mathbb{O}}$ can be estimated by a score based
approach as follows. For each set $S\subseteq\{1,\dots,d\}$ compute
two scores: (1) A stability score, denoted by $\stabscore(S)$, which
measures how well the regression based on predictors from $S$
satisfies the invariance \eqref{eq:invariant_mechanism_assumption} and
(2) a prediction score, denoted by $\predscore(S)$, which measures how
predictive the regression based on predictors from $S$ is. Based on
these scores, estimate the collection of generalizable sets as
\begin{equation*}
  \widehat{\mathbb{G}}\coloneqq\big\{S\subseteq\{1,\dots,d\}\,\big\vert\,
    \stabscore(S)\geq c_{\operatorname{stab}}\big\}
\end{equation*}
and the collection of generalizable and regression optimal sets as
\begin{equation*}
  \widehat{\mathbb{O}}\coloneqq\big\{S\in\widehat{\mathbb{G}}\,\big\vert\,
  \predscore(S)\geq c_{\operatorname{pred}}\big\}.  
\end{equation*}
The cutoff parameters $c_{\operatorname{stab}}$ and
$c_{\operatorname{pred}}$ are tuning parameters. Depending on the
data, the regression technique and potential domain knowledge,
different types of scores and cutoffs can be selected.

Below, we discuss several explicit options for constructing stability
and prediction scores. Here, we focus on settings where the response
can be expressed as a function of the predictors with additive noise,
i.e., $Y=f(X) + \epsilon$. For the stability score, we suggest using
an approximate hypothesis test for the null hypothesis
$S\in\mathbb{G}_{\mathcal{E}^{\text{obs}}}$ (see
Section~\ref{sec:stabscore}). For the prediction score, a bootstrap
approach based on mean squared errors can be employed (see
Section~\ref{sec:predscore})

\subsubsection{Stability scores}\label{sec:stabscore}
We propose to construct stability scores for each set
$S\subseteq\{1,\dots,d\}$ by a hypothesis test for the null hypothesis
$S\in\mathbb{G}_{\mathcal{E}^{\text{obs}}}$, i.e., whether $S$
satisfies the invariance
\eqref{eq:invariant_mechanism_assumption}. Once such a test has been
selected, we set, for any set $S\subseteq\{1,\dots,d\}$, the stability
score $\stabscore(S)$ to be the p-value of this test. An intuitive
parameterization is to set the cutoff $c_{\operatorname{stab}}$ to be
the type-1 error control for the hypothesis test, which controls the
trade-off of how stringently we want to enforce stability.

There are many ways in which a hypothesis test for this problem can be
constructed. Here, we discuss some potential starting points for the
general case and conclude with two well-known tests for Gaussian
linear models. Assume we fit a regression function $\hat{f}^{S}_{e}$
on each observed environment $e\in\mathcal{E}^{\text{obs}}$
individually. Given the null hypothesis
$S\in\mathbb{G}_{\mathcal{E}^{\text{obs}}}$, all of these regression
functions should be approximately equal up to the error from the
estimation, i.e., $\hat{f}^{S}_{e} \approx \hat{f}^{S}_{h}$. As a
consequence, the residuals $\hat{R}_e^S=Y_e-\hat{f}^{S}_{e}(X_e^S)$ on
each environment should also have approximately the same distribution,
i.e., $\hat{R}^{S}_{e} \overset{d}{\approx} \hat{R}^{S}_{h}$. One can
therefore construct a hypothesis test explicitly quantifying the
estimation error in either of these approximations. However, in order
to be able to do this, one needs to make some assumptions on the data
generating process.

In the case of linear regression, when the data generating process is
a linear model with Gaussian noise ($Y=\beta X + \epsilon$), we can
explicitly test for equal regression parameters $\hat{\beta}_e$ and
$\hat{\beta}_h$ using a Chow test \citep{chow1960}. A slight
disadvantage of this test is that it can only test equivalence between
two environments at a time. This means one needs to correct for
multiple testing whenever there are more than two environments. A
second option in the Gaussian linear case is to use a resampling based
test as suggested by \citep{shah2015}. One can show that it is
possible to exactly resample from the distribution of the scaled
residuals $R_e/\norm{R_e}_2$. This allows to construct a test for an
arbitrary test statistic based on $R_e/\norm{R_e}_2$ (e.g., the sum of
differences in mean across environments).

\subsubsection{Prediction scores}\label{sec:predscore}
For the prediction score, we propose to either use the negative mean
squared prediction error or the negative minimal environment-wise mean
squared prediction error. We use negative values to ensure that large
values imply predictive and small values non-predictive. To make the
cutoff interpretable and easier to select, one can use the following
bootstrap procedure.  For every set $S\subseteq\{1,\dots,d\}$, let
$\predscore(S)$ be the chosen prediction score. Construct $B$
bootstrap samples, $(\vX_1^{*},\vY_1^*),\dots,(\vX_B^{*},\vY_B^*)$,
and define for every $S\subseteq\{1,\dots,d\}$ the bootstrap
distribution function of the prediction score for all $t\in\R$ as
\begin{equation*}
  F_{\predscore(S)}^{*}(t)\coloneqq
  \sum_{i=1}^{B}\mathds{1}_{\{\predscore(S)(\vX_i^{*},\vY_i^{*})\leq t\}}.
\end{equation*}
Moreover, let $Q\in\widehat{\mathbb{G}}$ be the set of predictors with
maximal prediction score, i.e.,
$Q\coloneqq\argmax_{S\in\widehat{\mathbb{G}}}\predscore(S)(\vX,\vY)$.
Then, we choose the cutoff parameter to be
\begin{equation*}
  c_{\operatorname{pred}}=(F_{\predscore(Q)}^{*})^{-1}(\alpha_{\operatorname{pred}}),
\end{equation*}
where $\alpha_{\operatorname{pred}}\in (0,1)$ specifies how strongly
one wants to focus on the best predicting set.

\subsection{Variable importance}\label{sec:vi_sr}

Based on the stabilized regression estimator it is possible to define
several types of variable importance measures that can then be used to
recover either the Markov blanket, the stable blanket or the
non-stable blanket.

Assume we have computed the stabilized regression estimator given in
\eqref{eq:stabilizedregression_estimator}. Then, for each variable
$j\in\{1,\dots,d\}$, define the weight variable importance as follows
\begin{equation*}
  v^{\text{weight}}_{j}\coloneqq\sum_{S\subseteq\{1,\dots,
    d\}}\widehat{w}_S\cdot\mathds{1}_{\{j\in S\}}.
\end{equation*}
This means, the importance of a variable depends on how often it
appears with a positive weight. In the case of linear regression a
similar importance measure can be defined. To that end, let the
individual regression functions be given by
$\widehat{f}_S: x\mapsto \widehat{\beta}_S^{\top}x$, where
$\widehat{\beta}_{S}$ is the ordinary least squares estimator based on
the predictor set $S$ with zeros at all other coordinates. Then,
define the coefficient variable importance as
\begin{equation*}
  v^{\text{coef}}_{j}\coloneqq 2^{-d}\sum_{S\subseteq\{1,\dots,
    d\}}\abs{\widehat{\beta}_S^j}.
\end{equation*}
A third option, that can be used for a general regression procedure,
is a permutation based approached. Let $\vX^{*,j}_1,\dots,\vX^{*,j}_B$
be permuted versions of the data in which the $j$-th coordinate is
permuted while the remaining coordinates are fixed. Then, the
permutation variable importance is defined as
\begin{equation*}
  v^{\text{perm}}_{j}\coloneqq \frac{1}{B}\sum_{i=1}^B\left(\frac{\text{RSS}^{*,j}_i-\text{RSS}}{\text{RSS}}\right),
\end{equation*}
where $\text{RSS}$ and $\text{RSS}^{*,j}_i$ are the residual sum of
squares of the estimator $\hat{f}_{\operatorname{SR}}$ based on the
training data $\vX$ and the permuted data $\vX^{*,j}_i$, respectively.

Since stabilized regression, averages over the sets that are estimated
to be generalizable and regression optimal, using any of these
variable importance measures should rank variables higher if they
belong $\mathcal{O}_{\mathcal{E}^{\text{obs}}}$. In relation to
Section~\ref{sec:SP_vars} we hope that variables in the stable blanket
are ranked higher. Similarly, if the stability test cutoff is removed
or, equivalently, set to $-\infty$, in stabilized regression, the
variables importance should rank variables higher that are in the
Markov blanket. A sensible ranking for whether a variable belongs to
the non-stable blanket is thus given by the variable importance
\begin{equation*}
  v^{\text{SRdiff}}_j\coloneqq v^{\text{SRpred}}_j-v^{\text{SR}}_j,
\end{equation*}
where $v^{\text{SR}}_j$ and $v^{\text{SRpred}}_j$ are one of the
variable rankings above, based on stabilized regression with and
without stability cutoff, respectively.

\subsection{Implementation}\label{sec:implementation}
Given a regression procedure, stabilized regression is straightforward
to implement and pseudo-code is given in Algorithm~\ref{alg:SR}. The
framework is modular and most components such as stability score,
prediction score, variable screening and subsampling of subsets can
all be adjusted according to the application at hand.

\begin{algorithm}[h]
\SetKwInOut{Input}{input}
\SetKwInOut{Output}{output}
\Input{predictor matrix $\vX$\newline
  response matrix $\vY$\newline
  environments $\mathcal{E}^{\text{obs}}$\newline
  parameters
  $\alpha_{\operatorname{pred}}, \alpha_{\operatorname{stab}}\in (0,1)$
 }

 perform variable screening (optional)

 select collection of sets $\{S_1,\dots,S_M\}$ (all or subsampled)
 
 \For{$k\in\{1,\dots,M\}$}{
   fit regression function $\hat{f}^{S_k}$

   compute stability score $\stabscore(S)$

   compute prediction score $\predscore(S)$
 }

 $\widehat{\mathbb{G}}\gets \big\{S\in\{S_1,\dots,S_M\}\,\big\vert\, \stabscore(S)\geq \alpha_{\operatorname{stab}}\big\}$

 $c_{\operatorname{pred}}\gets(F_{\operatorname{MSE}_Q}^{*})^{-1}(1-\alpha_{\operatorname{pred}})$
 
 $\widehat{\mathbb{O}}\gets
 \big\{S\in\widehat{\mathbb{G}}\,\big\vert\, \predscore(S)\geq
 c_{\operatorname{pred}}\big\}$

 compute weights $\widehat{w}_S$ according to \eqref{eq:weights}
 
 \Output{weights $\widehat{w}_S$\newline
   regressors $\hat{f}^S$}
 \caption{StabilizedRegression}
 \label{alg:SR}
\end{algorithm}

In Algorithm~\ref{alg:SR}, we added a variable screening step in
line~1, since exhaustive subset search becomes infeasible as soon as
more than about $15$ variables are involved. Instead, we propose to
combine a variable screening with subsequent subsampling of predictor
sets. Any type of variable screening can be employed, as long as it
focuses on selecting predictive variables and removing irrelevant
variables. In the linear case, two reasonable approaches would be
either plain correlation screening \citep{fan2008sure} or an
$\ell^1$-penalty type screening as for example used in the Lasso
\citep{Tibshirani94}. How many variables to keep after screening
depends on the application. In general, our empirical analysis
suggested to screen as much as possible without removing any
potentially relevant predictors. To make computations feasible after
screening, one can additionally subsample subsets randomly. There are
several ideas that appear to work well in practice. Firstly, only
sample random sets up to a certain size. Ideally, if one has an
intuition about how many variables are required to get a stable and
predictive set (this can often be checked empirically) it makes sense
to put more weight on this expected size and only sample fewer
variables from smaller or larger set sizes. Secondly, the number of
subsampled sets should depend both on the expected number of stable
and predictive sets and on the number of variables after
screening. Empirically, it was often sufficient to subsample about
$1000$ sets, but generally the number should be selected in a data
driven fashion, similar to how the number of trees in a random forest
\citep{breiman2001random} is selected.

In Appendix~\ref{sec:defaults}, we make a detailed proposal on how to
choose default parameters.

\section{Numerical simulations}\label{sec:numerical_sim}

In this section, we assess the empirical performance of stabilized
regression. We restrict ourselves to the linear model setting, as this
is the setting of our biological application. First, in
Section~\ref{sec:sim_linear} we consider low dimensional linear
regression and in Section~\ref{sec:sim_highdim} high-dimensional
sparse linear regression. In both cases we assess how well stabilized
regression recovers the sets $\SB(Y)$ and $\NSB(Y)$ as well as the
predictive performance on unseen new environments.

\paragraph{Stabilized regression} Throughout this section, we use the
implementation of stabilized regression given in
Algorithm~\ref{alg:SR}. More specifically, we consider two versions
both using ordinary least squares as regression, but based on
different choices of weights $\widehat{w}_S$.  First, we use a vanilla
version denoted by SR. It uses the mean squared error as prediction
score and a resampling test based scaled residuals with the sum of
differences of environment-wise means as test statistic as stability
test (see Section~\ref{sec:stabscore}). The tuning parameters
$\alpha_{\text{pred}}$ and $\alpha_{\text{stab}}$ are both selected to
be $0.01$. Secondly, we use a predictive version, denoted by
SRpred. It uses the lowest environment-wise mean squared error as
prediction score (again with $\alpha_{\text{pred}}=0.01$) and does not
include any type of stability score. For both methods, we rank the
variables according to the score $v^{\text{coef}}_j$ defined in
Section~\ref{sec:vi_sr}. By construction, we expect SR to rank
variables in the stable blanket highest, while SRpred should rank
variables in the Markov blanket highest (as long as they are
predictive in at least one environment). We combine both procedures to
get a further variable ranking, denoted by SRdiff which ranks
variables according to
$v_j^{\text{SRdiff}}=v_j^{\text{SRpred}}-v_j^{\text{SR}}$ defined in
Section~\ref{sec:vi_sr}. We expect that this will recover variables in
the non-stable blanket. For the high-dimensional example, we combine
both stabilized regression procedures with $\ell^1$ pre-screening and
screen to $10$ variables.

\paragraph{Competing methods} As our simulations are all focused on
the linear case, we consider the following linear comparison methods:
(i) \emph{Ordinary linear least squares}. This method can only be
applied in the low-dimensional setting and will be denoted by
OLS. (ii) \emph{$\ell^1$-penalized linear regression}, also known as
Lasso \citep{Tibshirani94}, is a regularized version of linear
regression that is often employed to high-dimensional problems. We
select the penalty parameter based on cross-validation and denote the
method by Lasso. (iii) \emph{Anchor regression} due to
\citep{rothenhausler2018anchor}, which explicitly incorporates
heterogeneity. We consider two versions, one for the low-dimensional
case based on OLS and one for the high-dimensional case based on
Lasso, denoted by AR and AR (Lasso), respectively. The tuning
parameter for both is based on an environment-wise
cross-validation. (iv) \emph{Instrumental variables regression}, which
allows to guard against arbitrary shift strengths. We compute it via
the anchor regression estimate based on a penalty parameter of
$\gamma=1000$. As for the case of anchor regression there will be two
versions based either on OLS or Lasso, denoted by IV and IV (Lasso),
respectively. For each method, we get a variable importance measure by
taking the scaled regression parameter. All methods, except IV, should
recover the Markov blanket. On the other hand, in the settings
considered in the simulations, IV should recover the stable blanket
(see Section~\ref{sec:identifiability}) given a sufficient sample size
and strong enough interventions.

\subsection{Low-dimensional linear regression}\label{sec:sim_linear}

In our first numerical experiment, we consider a standard
low-dimensional linear SCM. We want to assess both the predictive
generalization performance as well as the variable selection. To this
end, we simulate $1000$ data sets according to
\href{dat:lowdim}{Simulation~1} and apply stabilized regression and
all competing methods to each.

\begin{figure}[h]
  \centering
  \resizebox{\textwidth}{!}{
    \includegraphics[width=\textwidth]{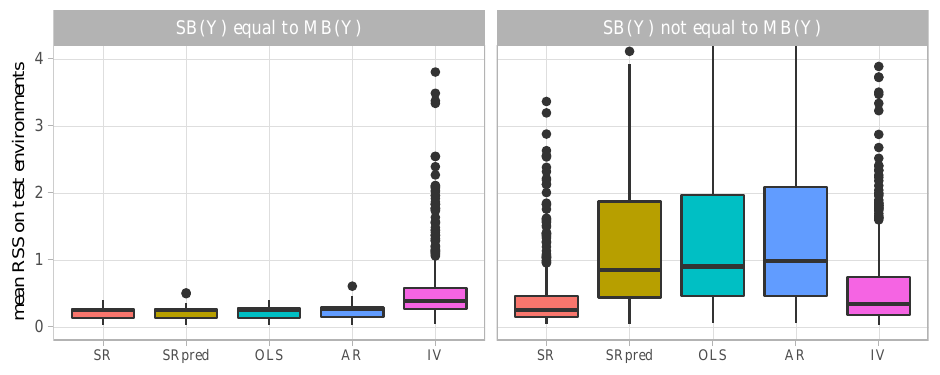}}
  \caption{Prediction results based on $1000$ repetitions from
    \href{dat:lowdim}{Simulation~1}. SR performs well both in the
    standard setting in which $\MB(Y)=\SB(Y)$ ($542$ repetitions) and
    the more difficult setting $\MB(Y)\neq\SB(Y)$ ($458$
    repetitions). Apart from SR and IV no other method is expected to
    generalize to these settings. The difference in performance
    between SR and IV is a finite sample property and shows that
    averaging can outperform direct optimization of the optimization
    \eqref{eq:constrained_opt}.}
  \label{fig:prediction_lowdim}
\end{figure}

\begin{mdframed}[roundcorner=5pt,
  frametitle={\hypertarget{dat:lowdim}{Simulation 1}: Low-dimensional
    linear regression}]
  Randomly sample a DAG with $d=11$ variables as follows: (i) Sample a
  causal ordering by randomly permuting the variables. (ii) Iterate
  over the variable and sample for each variable at most $4$ parents
  from all variables with higher causal ordering.  Next, select a
  random node to be the response $Y$ and extend the DAG by randomly
  sampling $4$ variables from the remaining $d-1$
  variables and add a parent intervention node $I$ to each of
  them. Denote the adjacency matrix of the resulting DAG by $B$, i.e.,
  $B_{i,j}\neq 0$ if and only if there is an edge from node $i$ to
  node $j$. For each non-zero entry in $B$ sample an edge weight
  uniformly from $(-1.5,-0.5)\cup(0.5, 1.5)$.

  Based on this DAG, generate data from different environments
  consisting of random mean shift in the noise of the intervention
  variables. The random mean shifts are sampled differently depending
  on whether the environment is used for training or for
  testing. Specifically, for training the mean shift is sampled
  uniformly from $(-1, 1)$ and for testing it is sampled uniformly from
  $(-10, 10)$. Based on these settings, sample $5$ training and
  $10$ testing environments each consisting of $n=250$ observations
  using Gaussian noise. More specifically, for each environment $e$
  generate data according to
  \begin{equation*}
    \vX_e = (\vI-B)^{-1}\bm{\epsilon}_e,
  \end{equation*}
  where $\bm{\epsilon}_e\in\R^{n\times (d+1)}$ and each row is sampled
  multivariate normal with covariance matrix $0.25\cdot\vI$ and mean
  vector $\mu$ which specifies the random mean shift for the
  intervention variables and is zero everywhere else.
\end{mdframed}
The prediction performance (in terms of mean residual sum of squares)
on the testing environments is given in
Figure~\ref{fig:prediction_lowdim}. The $1000$ repetitions are split
depending on whether $\MB(Y)=\SB(Y)$ or $\MB(Y)\neq\SB(Y)$ ($542$
repetitions in the first and $458$ repetitions in the second case). In
the case that $\MB(Y)=\SB(Y)$, we expect all procedures to perform
similarly as all prediction method should be generalizable in this
case. Only the IV method performs slightly worse, which is expected
since it generally is an estimator with higher variance. On the other
hand, in the case $\MB(Y)\neq\SB(Y)$ not all methods generalize to the
training method. Only SR and IV are expected to be generalizable in
this case. However, IV again performs worse than SR. The reason that
AR does not generalize in this case is that the testing shifts are
chosen to be stronger than the training environments. It therefore is
not able to guard against these types of shifts.

\begin{figure}[h]
  \centering
  \resizebox{\textwidth}{!}{
    \includegraphics[width=\textwidth]{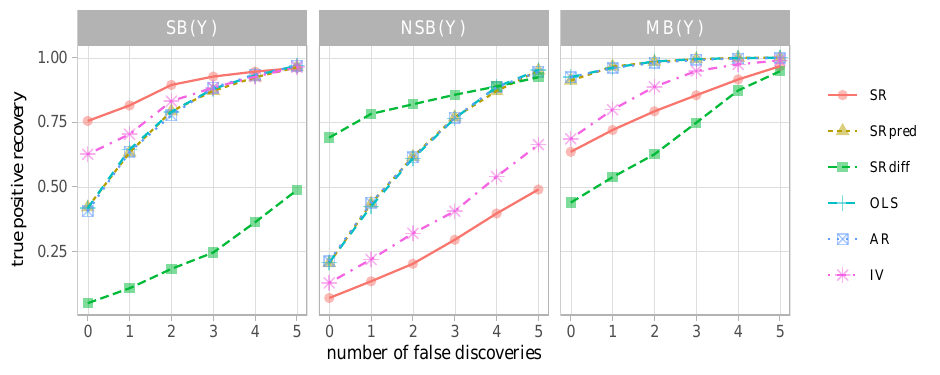}}
  \caption{Recovery performance based on $386$ repetitions (only using
    repetitions with $\SB(Y)\neq\varnothing$ and
    $\NSB(Y)\neq\varnothing$) from
    \href{dat:lowdim}{Simulation~1}. Each of the different versions of
    stabilized regression recovers one set well: SR has the best
    recovery of $\SB(Y)$, SRdiff has the best recovery of $\NSB(Y)$
    and SRpred performs competitive in recovering $\MB(Y)$.}
  \label{fig:vs_lowdim}
\end{figure}

Based on \href{dat:lowdim}{Simulation~1}, we can compute the ground
truth sets $\MB(Y)$, $\SB(Y)$ and $\NSB(Y)$ and check how well each
method recovers each of these sets. To this end, we compute true and
false positive rates for each method based on its variable importance
ranking. Results are given in Figure~\ref{fig:vs_lowdim}, where we
only consider the $386$ cases of the $1000$ repetitions for which
$\SB(Y)\neq\varnothing$ and $\NSB(Y)\neq\varnothing$. As one would
expect from the prediction results, SR outperforms the other methods
in terms of recovering the stable blanket. Since SR down-weights
variables in the non-stable blanket it performs poorly for recovering
$\MB(Y)$ and $\NSB(Y)$. However, SRpred is better in recovering the
Markov blanket (comparable with OLS) and hence SRdiff allows good
recovery of the $\NSB(Y)$. As expected AR and OLS both are good at
recovering $\MB(Y)$. However, they perform bad in terms of recovery of
both $\SB(Y)$ and $\NSB(Y)$ and hence themselves do not allow to
distinguish between them. IV on the other hand, solves the same
optimization as SR and hence aims at recovering $\SB(Y)$. Similarly,
it therefore also down-ranks variables from $\NSB(Y)$, but is not
quite as good as SR in this respect.

\FloatBarrier

\subsection{High-dimensional linear regression}\label{sec:sim_highdim}

\begin{figure}[h]
  \centering
  \resizebox{\textwidth}{!}{
    \includegraphics[width=\textwidth]{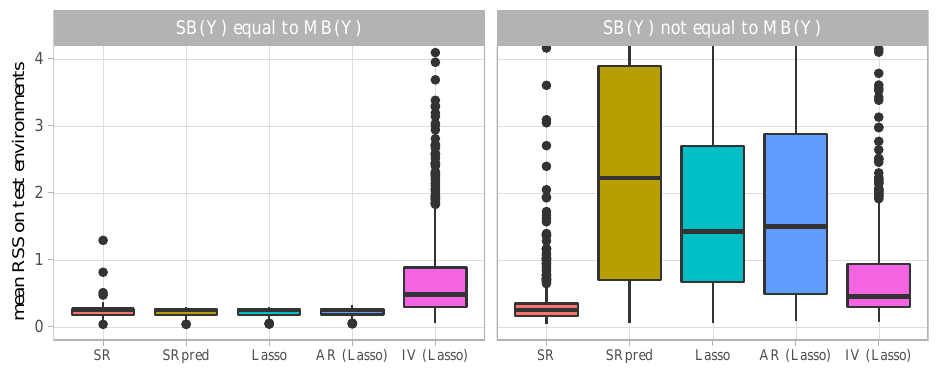}}
  \caption{Prediction results based on $1000$ repetitions from
    \href{dat:highdim}{Simulation~2}. SR performs well both in the
    standard setting in which $\MB(Y)=\SB(Y)$ ($643$ repetitions) and
    the more difficult setting $\MB(Y)\neq\SB(Y)$ ($357$
    repetitions). Apart from IV no other method is expected to
    generalize to these settings. The performance difference between
    IV and SR is even more pronounced in the high-dimensional
    settings.}
  \label{fig:prediction_sparse}
\end{figure}

\begin{figure}[h]
  \centering
  \resizebox{\textwidth}{!}{
    \includegraphics[width=\textwidth]{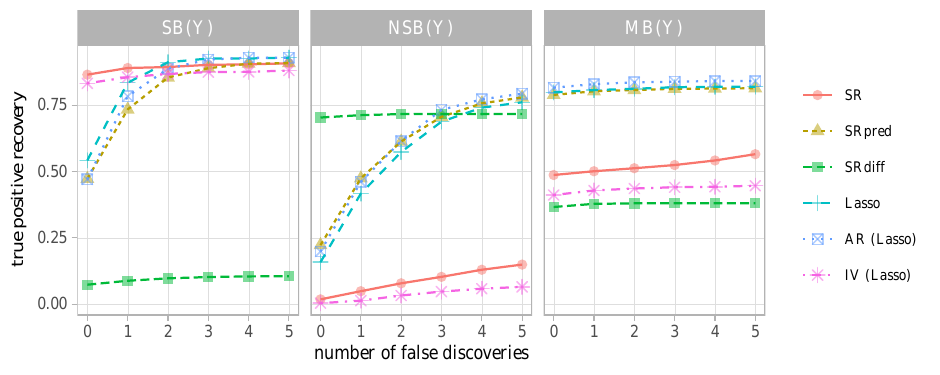}}
  \caption{Recovery performance based on $248$ repetitions (only using
    repetitions with $\SB(Y)\neq\varnothing$ and
    $\NSB(Y)\neq\varnothing$) from
    \href{dat:highdim}{Simulation~2}. Each of the different versions
    of stabilized regression recovers one set well: SR has the best
    recovery of $\SB(Y)$, SRdiff has the best recovery of $\NSB(Y)$
    and SRpred performs competitive in recovering $\MB(Y)$.}
  \label{fig:vs_sparse}
\end{figure}

To illustrate that stabilized regression adapts to high-dimensional
settings, we consider a high-dimensional linear setting. The data is
simulated according to \href{dat:highdim}{Simulation~2}. Again we
consider both prediction and variable selection properties of all
methods. Results are given in Figure~\ref{fig:prediction_sparse} and
Figure~\ref{fig:vs_sparse}. Overall, the results substantiate the
conclusions drawn in Section~\ref{sec:sim_linear}.

\begin{mdframed}[roundcorner=5pt,
  frametitle={\hypertarget{dat:highdim}{Simulation 2}: High-dimensional
    linear regression}]
  Randomly sample a DAG with $d=1001$ variables as follows: (i) Sample
  a causal ordering by randomly permuting the variables. (ii) From the
  full graph based on this causal order select include each edge with
  a probability of $p=2/(d-1)$, so the expected number of edges is
  $d$. Fix the first variable to be the response $Y$ and denote the
  adjacency matrix of the resulting DAG by $B$, i.e., $B_{i,j}\neq 0$
  if and only if there is an edge from node $i$ to node $j$. For each
  non-zero entry in $B$ sample an edge weight uniformly from
  $(-1.5,-0.5)\cup(0.5, 1.5)$.

  Based on this DAG, generate data from different environments
  consisting of random mean shift interventions on a random subset of
  the children of $Y$, which is selected by randomly choosing each
  child with probability $q=0.9$. The random mean shifts are sampled
  differently depending on whether the environment is used for
  training or for testing. Specifically, for training the mean shift
  is sampled uniformly from $(-1, 1)$ and for testing it sampled
  uniform from $(-10, 10)$. Based on these settings, sample $5$
  training and $10$ testing environments each consisting of $n=100$
  observations using Gaussian noise. More specifically, for each
  environment $e$ generate data according to
  \begin{equation*}
    \vX_e = (\vI-B)^{-1}\bm{\epsilon}_e,
  \end{equation*}
  where $\bm{\epsilon}_e\in\R^{n\times (d+1)}$ and each row is sampled
  multivariate normal with covariance matrix $0.25\cdot\vI$ and mean
  vector $\mu$ which specifies the random mean shift for the
  children that are intervened on and is zero everywhere else.
\end{mdframed}

\FloatBarrier

\section{Application to biological pathway analysis}\label{sec:application_bio}

In our application, we aim to generate novel biological hypotheses
about gene function. More specifically, we are interested in two types
of questions: (1) If we examine canonical metabolic pathways, can we
identify novel gene relationships interacting with the known pathway;
and (2) can we classify gene targets by whether they have a fixed or
switching functional dependence on a pathway's activity depending on
the environment. To answer these questions, we propose applying two
versions of stabilized regression and visualizing the results as in
Figure~\ref{fig:pathway_example}. The following steps describe the
procedure.
\begin{compactenum}
\item \textbf{Input:} A response variable $Y$ representing a quantity
  of interest (e.g., average activation levels of a pathway), a
  collection of gene expression levels $X^1,\dots,X^d$ and an
  environment variable $E$ indicating different conditions in which
  the data have been recorded.
\item \textbf{Stabilized regression:} Compute the following two
  versions of stabilized regression.
  \begin{compactenum}
  \item SR: Use the p-value of a stability test as stability score and
    pooled mean squared prediction error as prediction score.
  \item SRpred: Use the minimum environment-wise mean squared
    prediction error as prediction score and no stability cutoff.
  \end{compactenum}
  In both cases, we propose a correlation pre-screening to screen to
  approximately $\min_e \frac{n_e}{2}$ variables and a sub-sampling of
  subsets of a fixed maximum size (see
  Section~\ref{sec:implementation}).
\item \textbf{Variable importance:} Based on these two versions of
  stabilized regression, compute variable importance scores
  $v^{\text{SR}}_j$, $v^{\text{SRpred}}_j$ and $v^{\text{SRdiff}}_j$,
  using one of the variable importance measures from
  Section~\ref{sec:vi_sr}.
\item \textbf{Stability selection:} Use stability selection
  \citep{meinshausen2010stability} to compute selection probabilities
  for the two selection criteria $v^{\text{SR}}_j>0$ and
  $v^{\text{SRdiff}}_j>0$. This introduces sample stability into
  the estimates, hence increasing reliability of the results.
\item \textbf{Visualization:} Plot the two types selection
  probabilities on different axes (x-axis: SRdiff, y-axis: SR).
\end{compactenum}
The resulting plot visualizes the relation of all predicting genes with
the response. It allows explicitly distinguishing between genes that
have a stable functional dependence with the response across all
environments and genes that are predictive but have a functional shift
with respect to the response across environments. The stability
selection procedure adds a theoretical guarantee on the false positive
rate, which can be selected by practitioners (green regions in the
plot correspond to the threshold at which the expected number of
wrongly selected variables is at most $1$).

In the following section, we assess how well these plots work on a
transcriptomic and proteomic data set generated in the livers from an
aging mouse population \citep{roy2019}. To this end, we first
benchmark our method with other common approaches used to find
functionally related genes (Section~\ref{sec:recovery}). Secondly, we
discuss whether our method is able to distinguish between stable and
unstable dependencies (Section~\ref{sec:stablegenes}). In all of the
following experiments, we use a stability score based on the Chow test
and set the cutoff parameters to $\alpha_{\text{pred}}=0.01$ and
$\alpha_{\text{stab}}=0.1$. Furthermore, we use correlation
pre-screening and screen to $50$ variables. Finally, we sub-sample
$5000$ subsets consisting of at most $6$ variables and use
$v^{\text{coef}}_j$ as variable importance measure.

\subsection{Gene recovery}\label{sec:recovery}

\begin{figure}[h]
  \centering
  \includegraphics[width=\textwidth]{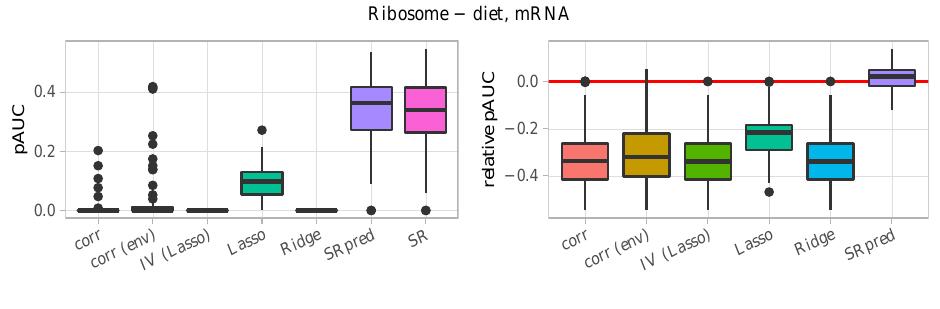}
  \caption{Recovery analysis with mRNA data based on the
    \emph{Ribosome} pathway using diet as an environment
    variable. (Left) Box plot of pAUC values for recovery of different
    genes belonging to the Ribsome pathway. (Right) Relative
    difference of each pAUC value compared to SR. Values below $0$
    imply worse pAUC compared to SR. Stabilized regression outperforms
    the competing methods.}
  \label{fig:ribo_pathway_recovery}
\end{figure}

Validation on real data is often difficult and can only be as good as
the ground truth known about the underlying system. Here, as a rough
approximation, we assume that genes belonging to the same canonical
metabolic pathways are functionally closer than genes not belonging to
the same pathway \citep{francesconi2008}. Furthermore, data-driven
network approaches to functional gene annotation have proven
successful in independent de novo reconstitution of functional gene
ontology sets which have been curated over decades through molecular
experimentation \citep{dutkowski2013}. This assumption is key to any
correlation-based discovery approach in biology and is known to be
particularly well satisfied in larger protein complexes
\citep{Roumeliotis2017}. Our validation is based on taking a set of
genes from known metabolic pathways, iteratively taking each of these
genes as a response $Y$ and then observing how many canonical genes
from the known pathway are recovered. We selected $7$ pathways for
this analysis taken from the KEGG database \citep{KEGG} and the
Reactome Pathway Knowledgebase \citep{Reactome}. More details are
given in Appendix~\ref{sec:pathway_details}. In our analysis we use
diet (low-fat vs. high-fat) as an environment variable.
\begin{figure}[h]
  \centering
  \includegraphics[width=\textwidth]{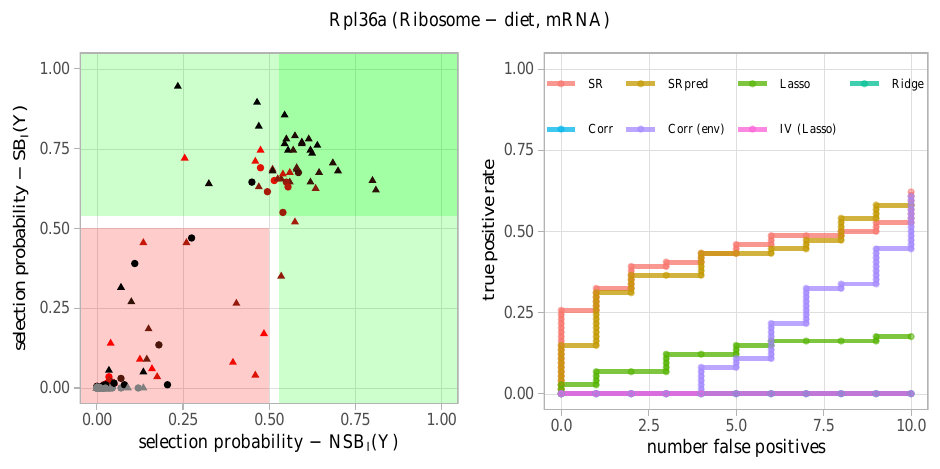}
  \caption{Recovery analysis for gene Rpl36a from the \emph{Ribosome}
    pathway. (Left) Visualization for Rpl36a as response and all
    remaining genes as potential predictors. Canonical \emph{Ribosome}
    genes are marked with a triangle all other genes with a
    circle. Many correct genes are ranked high. (Right) pROC for
    different methods, where canonical \emph{Ribosome} genes are
    considered true positives and all other genes false positives.}
  \label{fig:rpl36a}
\end{figure}
The result of applying the procedure described above to a single gene
from the \emph{Ribosome} pathway results in Figure~\ref{fig:rpl36a}
(left) -- Figure~\ref{fig:pathway_example} shows the same analysis for
a different pathway. To visualize which other genes belong to this
pathway, we have drawn these genes as triangles. We compare the
recovery performance of our method with other commonly applied methods
by computing partial receiver operator curves (pROC) with up to $10$
false positives as shown in Figure~\ref{fig:rpl36a} (right). Finally,
we did this for all genes from the pathway and summarized the
resulting pROCs using the normalized area under these curves, called
pAUC (partial area under the receiver operator curve). The results for
the \emph{Ribosome} pathway are shown in
Figure~\ref{fig:ribo_pathway_recovery}.

The results for all $7$ pathways (both for mRNA and protein data) are
given in Appendix~\ref{sec:bio_results}. While in many cases the
results are not as pronounced as for the Ribsome pathway, one can see
that in most cases stabilized regression performs at least as good as
other competitors and often better. The differences between methods is
less obvious for protein data, for which basic correlation screening
often performs very well. We believe this might be due to the fact
that proteins are one step closer to the biological processes and
hence these measurements capture the functional relations more
directly.

\subsection{The advantage of stability}\label{sec:stablegenes}

A key advantage of our method is that it allows to group genes based
on whether their dependence on the response is stable or unstable. We
illustrate this with Figure~\ref{fig:pathway_example} (and
Figure~\ref{fig:rpl36a} (left)). The green region of significant
findings can be divided into three parts, that should be interpreted
differently. The first region is the top left area of the plot. Genes
that appear there are detected only by SR and not by SRpred which
implies that they might not be the most predictive genes but depend on
the response in a stable fashion across all environments. The second
region is the bottom right part of the plot. These genes are only
found by SRpred and not by SR. This means that they are strongly
predictive for the response in at least one of the environments but
the dependence with the response changes across environments. Finally,
the third area is the top right corner of the plot, in which the green
areas overlap. Genes in this area are significantly reduced in
importance in SR compared to SRpred but still remain significant in
terms of SR. This can happen if the stability cutoff is not
consistently removing the same genes in all cases, which means that
the variations across environments are not strong enough in the data
to distinguish whether these genes are stable or unstable. While no
conclusion can be drawn on whether these genes are stable or unstable,
they can be considered to be predictive for the response.

\section{Discussion}

We propose a regression framework for multi-environment settings. Our
novel algorithm, stabilized regression, averages over regression
estimates based on subsets of predictors, in order to regularize the
final predictions to be both predictive and stable across
environments. We relate this setting to causal models and prove that,
under mild conditions, there exists an optimal subset of predictors
called the stable blanket, which generalizes across environments,
while minimizing the mean squared prediction loss. Furthermore, we
show that one can separate the Markov blanket into the stable blanket
and the non-stable blanket, which allows to characterize predictive
variables by whether they have a stable or unstable functional
dependence on the response. Using this framework, we propose a
procedure that aides hypothesis generation in systems biology and
demonstrate its usefulness on a current multiomic data set. The
procedure is shown to perform well in terms of recovery on known
biological pathways and additionally allows to separate findings into
stable and unstable predictors.

While our framework can be combined with any regression procedure, we
focus on the case of linear models. Future research should therefore
assess how these ideas perform on nonlinear regression problems. In
those settings, one needs to be more careful about how to deal with
shift interventions, since extrapolation might not be well-defined
anymore. A further interesting path forward, would be to consider
different notions of stability other than the one considered here
based on the conditional invariance defined in
\eqref{eq:invariant_mechanism_assumption}.

%%% Local Variables:
%%% mode: latex
%%% TeX-master: "StabilizedRegression"
%%% End:

\FloatBarrier

%% Acknowledgements
\section*{Acknowledgements}
We thank Yuansi Chen and Nicolai Meinshausen for helpful
discussions. NP and PB were supported by the European Research
Commission grant 786461 CausalStats - ERC-2017-ADG. JP was supported
by a research grant (18968) from VILLUM FONDEN.

% Bibliography
\bibliographystyle{plainnat}
\bibliography{references}

% Appendix
\appendix
\section{Proofs}\label{sec:proofs}

\begin{proof}[Proposition~\ref{thm:invariance_equiv}]
  First, fix a new environment $e^*\in\mathcal{E}$. We will show that
  one can generalize to this environment. To this end, introduce an
  auxiliary random variable $E$ taking values in
  $\mathcal{E}^{\text{obs}}\cup\{e^*\}$ with equal probability (only
  for simplicity). This variable is used to model the environments. To
  do this, we construct a SCM $\mathcal{S}_{\text{full}}$ over the
  variables $(E, I, X, Y)$ by setting $(I, X, Y)$ to the assignments
  in $\mathcal{S}^{*}$. Since the interventions only change the
  structural assignments of $I$ the variables $E$ are source nodes
  with edges only to the variables $I$ in the induced graph
  $\mathcal{G}(\mathcal{S}_{\text{full}})$. In particular, the SCM
  $\mathcal{S}_{\text{full}}$ induces a distribution $P_{\text{full}}$
  which by the assumptions in Setting~\ref{setting:causal} has a
  density $p$ which factorizes with respect to a product
  measure. Moreover, since distributions induced by SCMs satisfy the
  Markov properties \citep{Pearl2009, Lauritzen1990} and since $S$ is
  intervention stable, it holds for $(E,I,X,Y)\sim P_{\text{full}}$
  that
  \begin{equation}
    \label{eq:cond_independence}
    E\independent Y \vert X^S.
  \end{equation}
  By Setting~\ref{setting:causal} it holds for every
  $e\in\mathcal{E}^{\text{obs}}\cup\{e^*\}$ that the distribution
  $P_{e}$ has a density $p_e$. It follows from the construction of
  $\mathcal{S}_{\text{full}}$ that for all
  $e\in\mathcal{E}^{\text{obs}}\cup\{e^*\}$ and for all
  $z\in\bm{\mathcal{I}}\times\bm{\mathcal{X}}\times\R$ it holds that
  \begin{equation}
    \label{eq:conditioning}
    p(z\vert E=e)=p_e(z).
  \end{equation}
  Using this equation, the properties of conditional densities and
  \eqref{eq:cond_independence} it holds for all
  $e\in\mathcal{E}^{\text{obs}}\cup\{e^*\}$,
  $x\in\bm{\mathcal{X}}^{\abs{S}}$ and $y\in\R$ that
  \begin{align*}
    p_e(y\vert X^{S}=x)
    &=p(y\vert E=e, X^s=x)\\
    &=\frac{p(y, e\vert X^s=x)}{p(e\vert X^s=x)}\\
    &=\frac{p(y\vert X^s=x)p(e\vert X^s=x)}{p(e\vert X^s=x)}\\
    &=p(y\vert X^s=x).
  \end{align*}
  Finally, we use these expressions to compute the conditional
  expectation functions. Let
  $e\in\mathcal{E}^{\text{obs}}\cup\{e^*\}$, then for all
  $x\in\bm{\mathcal{X}}^{\abs{S}}$ it holds that
  \begin{align*}
    \E\left(Y_e\vert X_e^{S}=x\right)
    &=\int_{\R}y p_{e}(y\vert X^s=x)dy\\
    &=\int_{\R}y p(y\vert X^s=x)dy,
  \end{align*}
  where the last expression does not depend on the value of $e$. Since
  $e^*\in\mathcal{E}$ was arbitrary, this completes the proof of
  Proposition~\ref{thm:invariance_equiv}.
\end{proof}

\begin{proof}[Theorem~\ref{thm:stable_blanket}]
  First, by the acyclic assumption in Setting~\ref{setting:causal}, it
  holds that $\PA(Y)\subseteq N^{\text{int}}$, which immediately also implies that
  $\PA(Y)\subseteq\SB(Y)$. We prove that $\SB(Y)\subseteq\MB(Y)$ by
  contradiction. Fix, $j\in\SB(Y)\setminus\MB(Y)$ and define
  $S^*\coloneqq\SB(Y)\setminus\{j\}$. By definition of the stable
  blanket this implies that
  \begin{equation*}
    \exists \bar{k}\in N^{\text{int}}\setminus S^*:\quad
    X^{\bar{k}}\nindependentG Y\,\vert\, X^{S^*}.
  \end{equation*}
  Note, that $\bar{k}\not\in\MB(Y)$ as this either means it would also
  be d-connected to $Y$ given $\SB(Y)$ or that $j\in\MB(Y)$, neither
  of which can be true. Fix a path from $X^{\bar{k}}$ to $Y$ that is
  d-connected given $S^*$. Since $\PA(Y)\subseteq S^*$, this path has
  to enter $Y$ via a child. There are two forms that this path can
  have:
  \begin{enumerate}[label=(\arabic*)]
  \item $X^{\bar{k}}\cdots X^{r}\rightarrow
    X^{l}\leftarrow Y$\label{item:pathone}
  \item $X^{\bar{k}}\cdots \leftarrow
    X^{l}\leftarrow Y$\label{item:pathtwo}
  \end{enumerate}
  For the path in case \ref{item:pathone} to be d-connected given
  $S^*$ we require that $r\in S^*$ and
  $\DE(X^{l})\cap S^*\neq\varnothing$. However, this also implies that
  $X^{\bar{k}}\nindependentG Y\,\vert\, X^{\SB(Y)}$ (since $r\neq
  j$). By definition of $\SB(Y)$ this implies that
  $r\not\in N^{\text{int}}$, which moreover implies that
  $\DE(X^{l})\cap \SB(Y)=\varnothing$ and hence is a contradiction. On
  the other hand, for path \ref{item:pathtwo} to be d-connected, we
  require that $l\not\in S^*$. Since $l\neq j$ this implies that
  $l\not\in\SB(Y)$, which moreover implies that
  $l\not\in N^{\text{int}}$ since $l\in\CH(Y)$. Hence, either
  $\bar{k}\in\DE(X^{l})$ which leads to a contradiction as this
  implies $\bar{k}\not\in N^{\text{int}}$ or there is a collider on
  the path. Let $X^{q}$ denote the collider closest to $X^l$. Then,
  since $q\in\DE(X^{l})$ (and hence $\DE(X^{q})\not\in N^{\text{int}}$)
  this collider implies that the path is not d-connected given $S^*$,
  which is a contradiction. This proves that $\SB(Y)\subseteq\MB(Y)$.

  Next, we show that $\SB(Y)$ is intervention stable, by proving that
  any path from an intervention variable to $Y$ is not d-connected
  given the stable blanket. Fix $\ell\in\{1,\dots,m\}$ and let $P$ be
  a path from $I^{\ell}$ to $Y$. First, note that any path entering
  $Y$ through a parent is blocked given $\SB(Y)$ (since
  $\PA(Y)\subseteq\SB(Y)$). So assume that $P$ enters $Y$ through
  $X^k$ with $k\in\CH(Y)$. There are two possible options for $P$:
  \begin{enumerate}[label=(\roman*)]
  \item $I^{\ell}\rightarrow\cdots X^{r}\rightarrow
    X^{k}\leftarrow Y$\label{item:path1}
  \item $I^{\ell}\rightarrow\cdots X^{r}\leftarrow
    X^{k}\leftarrow Y$\label{item:path2}
  \end{enumerate}
  In order for the path in option \ref{item:path1} to be d-connected
  given $\SB(Y)$, we need $r\not\in\SB(Y)$ and
  $\DE(X^k)\cap\SB(Y)\neq\varnothing$. There are exactly two cases in
  which $r\not\in\SB(Y)$: First, $r\in N^{\text{int}}$ in which case
  $X^r\independentG Y\,\vert\, X^{\SB(Y)}$ implying that the path is
  blocked. Second, $r\not\in N^{\text{int}}$ which by the definition
  of $N^{\text{int}}$ implies that
  $\DE(X^r)\cap N^{\text{int}}=\varnothing$. Since
  $\DE(X^k)\subseteq\DE(X^r)$ this, however, implies that
  $\DE(X^k)\cap\SB(Y)\neq\varnothing$. Path \ref{item:path1} can
  hence not be d-connected given $\SB(Y)$. In the case of option
  \ref{item:path2}, let $X^q$ be the collider on $P$ that is closest
  to $X^r$. This path is only d-connected given $\SB(Y)$ if
  $k\not\in\SB(Y)$ and $\DE(X^q)\cap\SB(Y)\neq\varnothing$. So assume
  that $k\not\in\SB(Y)$ then it holds that $k\not\in N^{\text{int}}$
  (otherwise $X^k\independentG Y\,\vert\, X^{\SB(Y)}$ which is
  impossible as $k\in\CH(Y)$). Now, since $\DE(X^q)\subseteq\DE(X^k)$
  and $\SB(Y)\subseteq N^{\text{int}}$, this implies that
  $\DE(X^q)\cap\SB(Y)=\varnothing$. Therefore, path \ref{item:path2}
  is not d-connected given $\SB(Y)$.

  This proves that $\SB(Y)$ is intervention stable. By
  Proposition~\ref{thm:invariance_equiv}, we hence know that $\SB(Y)$
  is generalizable. It therefore remains to prove that it is also
  regression optimal. To this end we use the following two properties
  of $N^{\text{int}}$:
  \begin{enumerate}[label=(\alph*)]
  \item For all $S\subseteq\{1,\dots,d\}$ being intervention stable it holds
    that $S\subseteq N^{\text{int}}$,\label{item:stat1}
  \item $N^{\text{int}}$ is intervention stable and hence
    generalizable.\label{item:stat2}
  \end{enumerate}
  Statement \ref{item:stat1} holds because conditioning on any
  $j\in\{1,\dots,d\}\setminus N^{\text{int}}$ immediately d-connects a
  path of the form $I^{\ell}\rightarrow X^k\leftarrow Y$, which cannot
  be blocked. The proof of statement \ref{item:stat2} is almost
  identical to the arguments above (using paths (i) and (ii)) and is
  not repeated here.

  Fix a set $\bar{S}\in\mathbb{O}_{\mathcal{E}}$ (it is non-empty
  since $\mathbb{G}_{\mathcal{E}}\neq\varnothing$). The orthogonality
  of the conditional expectation \citep[chapter 5]{Kallenberg} states
  that for all $Z\in L^{2}(\Omega, \sigma(X^{N^{\text{int}}}), \P)$ it holds that
  \begin{equation}
    \label{eq:cond_ortho}
    \E\big((Y-Z)^2\big)\geq\E\big((Y-\E(Y\vert X^{N^{\text{int}}}))^2\big).
  \end{equation}
  By statement \ref{item:stat1} it holds that
  $\bar{S}\subseteq N^{\text{int}}$, which implies that
  $\sigma(X^{\bar{S}})\subseteq\sigma(X^{N^{\text{int}}})$ and hence
  $L^{2}(\Omega, \sigma(X^{\bar{S}}), \P)\subseteq L^{2}(\Omega,
  \sigma(X^{N^{\text{int}}}), \P)$ (where $\sigma(\cdot)$ denotes the
  generated sigma algebra). Therefore, using \eqref{eq:cond_ortho}
  together with the fact that $\bar{S}\in\mathbb{O}_{\mathcal{E}}$, we
  get that
  \begin{equation}
    \label{eq:Sstar_pred}
    v_{\min}\coloneqq\E\big((Y-\E(Y\vert X^{\bar{S}}))^2\big)\geq\E\big((Y-\E(Y\vert X^{N^{\text{int}}}))^2\big),
  \end{equation}
  which together with statement \ref{item:stat2} implies that
  $N^{\text{int}}\in\mathbb{O}_{\mathcal{E}}$. By the definition of
  the stable blanket it holds for all
  $j\in N^{\text{int}}\setminus\SB(Y)$ that
  \begin{equation}
    \label{eq:conditional_ind_SB}
    X^j\independentG Y \,\vert\, X^{\SB(Y)}.
  \end{equation}
  Combining \eqref{eq:conditional_ind_SB} with Doob's conditional
  independence property \citep[Proposition~5.6]{Kallenberg} and
  \eqref{eq:Sstar_pred}, we get that
  \begin{align*}
    v_{\min}
    &\geq\E\big((Y-\E(Y\vert X^{N^{\text{int}}}))^2\big)\\
    &=\E\big((Y-\E(Y\vert X^{N^{\text{int}}\setminus\SB(Y)}, X^{\SB(Y)}))^2\big)\\
    &=\E\big((Y-\E(Y\vert X^{\SB(Y)}))^2\big),
  \end{align*}
  which proves that $\SB(Y)\in\mathbb{O}_{\mathcal{E}}$. This
  completes the proof of Theorem~\ref{thm:stable_blanket}.
\end{proof}

\begin{proof}[Lemma~\ref{thm:OLS_SCM}]
  Since SCMs are by definition acyclic it holds that the matrix $(\vI
  - B)$ is invertible and
  \begin{equation*}
    \begin{pmatrix}
      Y\\X
    \end{pmatrix}
    =(\vI-B)^{-1}\epsilon.
  \end{equation*}
  Hence, for $P\coloneqq(0, \vI)\in\R^{d\times (d+1)}$ it holds that
  \begin{equation}
    \label{eq:decompX}
    X=P(\vI-B)^{-1}\epsilon.
  \end{equation}
  The population ordinary least squares is given by
  \begin{equation*}
    \beta^{\operatorname{OLS}}=\cov(X)^{-1}\cov(X, Y).
  \end{equation*}
  Together with the expression $Y=\beta_{\tPA}^{\top}X+\epsilon^0$,
  which is a consequence of the SCM, it holds that
  \begin{align}
    \beta^{\operatorname{OLS}}
    &=\cov(X)^{-1}\left(\cov(X)\beta_{\tPA}+\cov(X,
      \epsilon^{0})\right)\nonumber \\
    &=\beta_{\tPA}+\cov(X)^{-1}\cov(X, \epsilon^{0})\nonumber\\
    &=\beta_{\tPA}+\left[P(\vI-B)^{-1}\cov(\epsilon)(\vI-B)^{-\top}P^{\top}\right]^{-1}P(\vI-B)^{-1}\cov(\epsilon, \epsilon^{0}).\label{eq:OLS_decomp}
  \end{align}
  Next, decompose the following three matrices
  \begin{equation*}
    (\vI-B)=
    \begin{pmatrix}
      1 &-\beta_{\tPA}^{\top}\\
      -\beta_{\tCH} &\vI-B_{X}
    \end{pmatrix},
    \quad\quad
    (\vI-B)^{-1}=
    \begin{pmatrix}
      1 &v^{\top}\\
      w &M
    \end{pmatrix},
    \quad\quad
    \cov(\epsilon)=
    \begin{pmatrix}
      \var(\epsilon^0) &0\\
      0 &D
    \end{pmatrix}.
  \end{equation*}
  Based on these decompositions, an explicit computation of
  \eqref{eq:OLS_decomp} leads to
  \begin{equation}
    \label{eq:OLS_decomp2}
    \beta^{\operatorname{OLS}}=\beta_{\tPA}+
    \left(\var(\epsilon^0)ww^{\top}+MDM^{\top}\right)^{-1}
    w
    \var(\epsilon^0).
  \end{equation}
  Furthermore, by the definition of the inverse we get that
  \begin{equation*}
    \vI=(\vI-B)(\vI-B)^{-1}=
    \begin{pmatrix}
      (1-\beta_{\tPA}^{\top}w) &v^{\top}-\beta_{\tPA}^{\top}M\\
      -\beta_{\tCH}+(\vI-B_{X})w &-\beta_{\tCH}v^{\top}+(\vI-B_{X})M\\
    \end{pmatrix}
  \end{equation*}
  and similarly that
  \begin{equation*}
    \vI=(\vI-B)^{-1}(\vI-B)=
    \begin{pmatrix}
      1-v^{\top}\beta_{\tCH} & -\beta_{\tPA}^{\top}+v^{\top}(\vI-B_{X})\\
      w-M\beta_{\tCH} &-w\beta_{\tPA}^{\top}+M(\vI-B_{X})\\
    \end{pmatrix}.
  \end{equation*}
  Solving these equations coordinate-wise leads to the following
  equalities, which we will refer to as \emph{inverse constraints},\\
  \begin{minipage}{0.5\textwidth}
    \begin{align*}
      \beta_{\tPA}^{\top}w&=0\\
      \beta_{\tCH}&=(\vI-B_{X})w\\
      v^{\top}&=\beta_{\tPA}^{\top}M\\
      \vI+\beta_{\tCH}v^{\top}&=(\vI-B_{X})M
    \end{align*}
  \end{minipage}
  \begin{minipage}{0.5\textwidth}
    \begin{align*}
      v^{\top}\beta_{\tCH}&=0\\
      w&=M\beta_{\tCH}\\
      \beta_{\tPA}^{\top}&=v^{\top}(\vI-B_{X})\\
      \vI+w\beta_{\tPA}^{\top}&=M(\vI-B_{X}).
    \end{align*}
  \end{minipage}
  
  \vspace{1em}
  \noindent Next, we use the Sherman–Morrison formula, given by
  \citet{bartlett1951}, to compute the inverse in
  \eqref{eq:OLS_decomp2} as follows
  \begin{align}
    \left(\var(\epsilon^0)ww^{\top}+MDM^{\top}\right)^{-1}
    = M^{-\top}D^{-1}M^{-1} -
      \frac{\var(\epsilon^{0})}{1+g}M^{-\top}D^{-1}M^{-1}ww^{\top}M^{-\top}D^{-1}M^{-1},
  \end{align}
  where
  \begin{equation*}
    g=\var(\epsilon^{0})w^{\top}M^{-\top}D^{-1}M^{-1}w.
  \end{equation*}
  This expression together with $M^{-1}w=\beta_{\tCH}$ (which
  follows from the inverse constraints) simplifies
  \eqref{eq:OLS_decomp2} as follows
  \begin{align}
    \label{eq:OLS_decomp3}
    \beta^{\operatorname{OLS}}
    &=\beta_{\tPA}+
      \left(M^{-\top}D^{-1}M^{-1} -
      \frac{\var(\epsilon^{0})}{1+g}M^{-\top}D^{-1}M^{-1}ww^{\top}M^{-\top}D^{-1}M^{-1}\right)w
      \var(\epsilon^0)\nonumber\\
    &=\beta_{\tPA}+
      M^{-\top}D^{-1}\beta_{\tCH}\var(\epsilon^0) -
      \frac{\var(\epsilon^{0})^2}{1+g}M^{-\top}D^{-1}\beta_{\tCH}\beta_{\tCH}^{\top}D^{-1}\beta_{\tCH}\nonumber\\
    &=\beta_{\tPA}+
      M^{-\top}D^{-1}\beta_{\tCH}\left(\var(\epsilon^0) -
      \frac{\var(\epsilon^{0})^2}{1+g}\beta_{\tCH}^{\top}D^{-1}\beta_{\tCH}\right).
  \end{align}
  The last step is to compute $M^{-\top}D^{-1}\beta_{\tCH}$. To this
  end, we use the inverse constraints to get that
  \begin{equation*}
    M^{-1}=\left((\vI-B_{X})^{-1}(\vI+\beta_{\tCH}v^{\top})\right)^{-1}
    =(\vI+\beta_{\tCH}v^{\top})^{-1}(\vI-B_{X}).
  \end{equation*}
  Again using the Sherman–Morrison formula and the inverse constraint
  $v^{\top}\beta_{\tCH}=0$ this implies that
  \begin{equation*}
    M^{-1}=(\vI-\beta_{\tCH}v^{\top})(\vI-B_{X}).
  \end{equation*}  
  Therefore, we get
  \begin{align*}
    M^{-\top}
    &=(\vI-B_{X})^{\top}(\vI-v\beta_{\tCH}^{\top})\\
    &=(\vI-B_{X})^{\top} -
      (\vI-B_{X})^{\top}v\beta_{\tCH}^{\top}\\
    &=(\vI-B_{X})^{\top} -
      \beta_{\tPA}\beta_{\tCH}^{\top},
  \end{align*}
  where in the last step we used the inverse constraints again.
  Finally, combining this with \eqref{eq:OLS_decomp3} and simplifying
  $g$ this leads to
  \begin{equation*}
    \beta^{\operatorname{OLS}}=\beta_{\tPA}+
      \left((\vI-B_{X})^{\top} - \beta_{\tPA}\beta_{\tCH}^{\top}\right)D^{-1}\beta_{\tCH}\left(1-
      \frac{\var(\epsilon^{0})\beta_{\tCH}^{\top}D^{-1}\beta_{\tCH}}{1+\var(\epsilon^0)\beta_{\tCH}^{\top}D^{-1}\beta_{\tCH}}\right)\var(\epsilon^0),
  \end{equation*}
  which completes the proof of Lemma~\ref{thm:OLS_SCM}.
\end{proof}

\begin{proof}[Corollary~\ref{thm:optimal_OLS}]
  The proof consists of an application of
  Lemma~\ref{thm:OLS_SCM}. In order to be able to apply the result,
  we fix $\bar{n}\in\N$ and construct a new SCM
  $\mathcal{S}^{\operatorname{tot}}_{\bar{n}}$ over the variables
  $(E_{\bar{n}}, \tilde{I}_{\bar{n}}, \tilde{X}_{\bar{n}}, \tilde{Y}_{\bar{n}})$, which includes both
  intervention SCMs. To this end, let $E_n$ be a Bernoulli random
  variable with probability $p=0.5$ and let
  $(\tilde{I}_{\bar{n}}, \tilde{X}_{\bar{n}}, \tilde{Y}_{\bar{n}})$ be such that it satisfies
  the SCM $\mathcal{S}_{e^{+}_{\bar{n}}}$ if $E_n=1$ and
  $\mathcal{S}_{e^{-}_{\bar{n}}}$ if $E_{\bar{n}}=0$. Furthermore, we get that
  \begin{align*}
    \E\left((\tilde{Y}_{\bar{n}}-\beta^{\top}\tilde{X}_{\bar{n}})^2\right)
    &=\frac{1}{2}\E\left((\tilde{Y}_{\bar{n}}-\beta^{\top}\tilde{X}_{\bar{n}})^2\,\vert\,
      E_{\bar{n}}=0\right)
      +\frac{1}{2}\E\left((\tilde{Y}_{\bar{n}}-\beta^{\top}\tilde{X}_{\bar{n}})^2\,\vert\,
      E_{\bar{n}}=1\right)\\
    &=\frac{1}{2}\sum_{e\in\mathcal{E}_{\bar{n}}}\E\left((Y_e-\beta^{\top}X_e)^2\right),
  \end{align*}
  where in the last step we used that by construction
  \begin{equation*}
    (\tilde{Y}_{\bar{n}},\tilde{X}_{\bar{n}})\,\vert\, (E_{\bar{n}}=0)
    \overset{d}{=}(Y_{e^{-}_{\bar{n}}},X_{e^{-}_{\bar{n}}})
    \quad\text{and}\quad
    (\tilde{Y}_{\bar{n}},\tilde{X}_{\bar{n}})\,\vert\, (E_{\bar{n}}=1)
    \overset{d}{=}(Y_{e^{+}_{\bar{n}}},X_{e^{+}_{\bar{n}}}).
  \end{equation*}
  This, in particular implies that the OLS of interest $\beta^{\operatorname{OLS}}_{\bar{n}}$
  is the same as the OLS resulting from regressing $\tilde{Y}_{\bar{n}}$ on
  $\tilde{X}_{\bar{n}}$.

  Next, we apply Lemma~\ref{thm:OLS_SCM} to the regression of
  $\tilde{Y}_{\bar{n}}$ on $\tilde{X}_{\bar{n}}$. To do this, we
  include the intervention variables into the noise variables of their
  respective children.  We denote the noise terms in this reduced SCM
  by $\epsilon^0_{\bar{n}},\dots,\epsilon^d_{\bar{n}}$, where
  $\epsilon^0_{\bar{n}}$ corresponds to $\tilde{Y}_{\bar{n}}$ and
  $\epsilon^k_{\bar{n}}$ to $\tilde{X}^k_{\bar{n}}$. By assumption,
  the intervention variables $I_{\bar{n}}$ each have exactly one child
  and are mutually independent. This implies that the noise matrix
  $D_{\bar{n}}=\cov(\epsilon^1_{\bar{n}},\dots,\epsilon^d_{\bar{n}})$
  is diagonal and that for all $k\in\{1,\dots,d\}$ it holds that
  $\var(I^{\ell}_{\bar{n}})=\sigma^2_{\bar{n}}$. From this point
  onward, we drop the notional dependence on $\bar{n}$ for
  convenience.

  Using Lemma~\ref{thm:OLS_SCM} and the notation therein, we get the
  following expression
  \begin{equation*}
    \beta^{\operatorname{OLS}}=\beta_{\tPA}+
    \left((\vI-B_{X})^{\top} - \beta_{\tPA}\beta_{\tCH}^{\top}\right)D^{-1}\beta_{\tCH}\left(1-
      \frac{\var(\epsilon^{0})\beta_{\tCH}^{\top}D^{-1}\beta_{\tCH}}{1+\var(\epsilon^0)\beta_{\tCH}^{\top}D^{-1}\beta_{\tCH}}\right)\var(\epsilon^0).
  \end{equation*}
  For all $j\in\{1,\dots,d\}$ this implies that
  \begin{equation}
    \label{eq:beta_minus_betaPA}
    \beta^{\operatorname{OLS},j}-\beta_{\tPA}^j=
    \sum_{k=1}^{d}\left((\vI-B_{X})_{k,j} - \beta_{\tPA}^{j}\beta_{\tCH}^{k}\right)\var(\epsilon^k)^{-1}\beta_{\tCH}^{k}K,
  \end{equation}
  with
  \begin{equation*}
    K=\left(1-\frac{\var(\epsilon^{0})\beta_{\tCH}^{\top}D^{-1}\beta_{\tCH}}{1+\var(\epsilon^0)\beta_{\tCH}^{\top}D^{-1}\beta_{\tCH}}\right)\var(\epsilon^0).
  \end{equation*}
  Now, let $j\in\{1,\dots,d\}\setminus\MB(Y)$, then
  $\beta_{\tPA}^{j}=0$. Moreover, for all $k\not\in\CH(Y)$ it holds
  that $\beta_{\tCH}^k=0$. Combining this with
  \eqref{eq:beta_minus_betaPA} implies
  \begin{equation*}
    \beta^{\operatorname{OLS},j}=
    \sum_{k\in\CH(Y)}-(B_{X})_{k,j} \beta_{\tCH}^{k}\var(\epsilon^k)^{-1}K,
  \end{equation*}
  where the $\vI$ term vanishes because $j \notin \CH(Y)$.  Since the
  SCM is assumed to be directed and acyclic it holds that $\MB(Y)$ is
  the union of $\PA(Y)$, $\CH(Y)$ and all $\PA(X^{\ell})$ for which
  $\ell\in\CH(Y)$. Hence, since $j\not\in\MB(Y)$, it holds that
  $\CH(Y)\cap\CH(X^j)=\varnothing$. Furthermore, because for all
  $k\not\in\CH(X^j)$ it holds that $(B_{X})_{k,j}=0$, this implies
  $\beta^{\operatorname{OLS},j}=0$.  It remains to show the result for
  $j\in\MB(Y)\setminus\SB(Y)$. To this end, we consider the following
  inequalities
  \begin{equation*}
    \abs{K}\leq \var(\epsilon^0)+\abs[\bigg]{\frac{1}{1 /
        \left(\var(\epsilon^0)\sum_{k=1}^d(\beta_{\tCH}^{k})^2\var(\epsilon^k)\right)+1}}\var(\epsilon^0)\leq 2\var(\epsilon^0).
  \end{equation*}
  It now suffices to show for all $j\in\MB(Y)\setminus\SB(Y)$ that
  \begin{equation}
    \label{eq:OLSdiff}
    \abs{\beta^{\operatorname{OLS},j}}\leq
    \sum_{k=1}^{d}\abs[\bigg]{\left((\vI-B_{X})_{k,j} -
        \beta_{\tPA}^{j}\beta_{\tCH}^{k}\right)\beta_{\tCH}^{k}}\cdot\frac{2\var(\epsilon^0)}{\var(\epsilon^k)}\longrightarrow 0,
  \end{equation}
  as $\bar{n}\rightarrow\infty$. Using again the decomposition of the Markov
  blanket and the fact that $\PA(Y)\subseteq\SB(Y)$, we need to
  consider the two possible cases (i) $j\in\CH(Y)$ and (ii) there
  exists $k\in\CH(Y)$ such that $j\in\PA(X^k)\setminus\CH(Y)$.
  
  For case (i): Let $j\in\CH(Y)$, then $j\not\in N^{\text{int}}$
  (since $j\not\in\SB(Y)$) and any set $S\subseteq\{1,\dots,d\}$ with
  $j\in S$ is not intervention stable. So in particular, there exists
  $\ell\in\{1,\dots,m\}$ such that
  \begin{equation*}
    I^{\ell}\nindependentG Y \,\vert\, X^{\{j\}\cup\PA(Y)}.
  \end{equation*}
  Therefore, there must be a d-connected path $P$ from $I^{\ell}$ to
  $Y$ given $X^{\{j\}\cup\PA(Y)}$ that enters $Y$ via $X^j$. Since we
  only condition on ${\{j\}}\cup\PA(Y)$ this implies that $X^j$ is the
  only collider and that $P$ contains a directed path from $I^{\ell}$
  to $X^j$. Now, since all directed paths are assumed to be
  non-vanishing, it holds for all $k\in\CH(X^j)\cup\{j\}$ that
  $\var(\epsilon^{k})\geq
  \text{const}\cdot\var(\tilde{I}^{\ell})=\mathcal{O}(\sigma_{\bar{n}}^2)$.
  Hence, using that $\beta_{\tPA}^j=0$, $\beta_{\tCH}^k=0$ for all
  $k\not\in\CH(Y)$ and $(B_{X})_{k,j}=0$ for all $k\not\in\CH(X^j)$ it
  follows from \eqref{eq:OLSdiff} that
  \begin{equation*}
    \lim_{\bar{n}\rightarrow\infty}\abs{\beta^{\operatorname{OLS},j}}\leq
    \lim_{\bar{n}\rightarrow\infty}\sum_{k\in\CH(Y)\cap\CH(X^j)}\abs[\big]{(\vI-B_{X})_{k,j}\beta_{\tCH}^{k}}\cdot\frac{2\var(\epsilon^0)}{\var(\epsilon^k)}=0.
  \end{equation*}

  For case (ii): Let $k\in\CH(Y)$ such that
  $j\in\PA(X^k)\setminus\CH(Y)$. We will show that for all
  $\bar{k}\in\CH(Y)\cap\CH(X^j)$ there exits $\ell\in\{1,\dots,m\}$
  such that for all $\bar{n}\in\N$ it holds that
  \begin{equation}
    \label{eq:req_bound}
    \var(\epsilon^{\bar{k}})\geq
    \text{const}\cdot\var(\tilde{I}^{\ell})=\mathcal{O}(\sigma_{\bar{n}}^2).
  \end{equation}
  To see this, fix $\bar{k}\in\CH(Y)\cap\CH(X^j)$, then we consider
  two cases (a) $\bar{k}\in\SB(Y)$ or (b) $\bar{k}\not\in\SB(Y)$. Case
  (a) leads to a contradiction: By definition of $\SB(Y)$, we get that
  $j\in N^{\text{int}}$. Moreover,
  $X^j\nindependentG Y\,\vert\, X^{\SB(Y)}$ because
  $\bar{k}\in\CH(Y)\cap\SB(Y)$. But this would mean that $j\in\SB(Y)$,
  which is a contradiction. Case (b) implies that
  $\bar{k}\not\in N^{\text{int}}$. Therefore, for all
  $S\subseteq\{1,\dots,d\}$ with $\bar{k}\in S$ the set $S$ is not
  intervention stable. So in particular, there exists
  $\ell\in\{1,\dots,m\}$ such that
  $I^{\ell}\nindependentG Y \,\vert\, X^{\PA(Y)\cup\{\bar{k}\}}$. As
  argued above this implies that there is a directed path from
  $I^{\ell}$ to $X^{\bar{k}}$ and since all directed paths are assumed
  to be non-vanishing, we get that
  $\var(\epsilon^{\bar{k}})\geq
  \text{const}\cdot\var(\tilde{I}^{\ell})=\mathcal{O}(\sigma_{\bar{n}}^2)$
  as desired. Finally, combining \eqref{eq:req_bound} with the fact
  that $\beta_{\tPA}^j=0$ and $\beta_{\tCH}^k=0$ for all
  $k\not\in\CH(Y)$ it follows from \eqref{eq:OLSdiff} that
  \begin{equation*}
    \lim_{\bar{n}\rightarrow\infty}\abs{\beta^{\operatorname{OLS},j}}\leq
    \lim_{\bar{n}\rightarrow\infty}\sum_{k\in\CH(Y)\cap\CH(X^j)}\abs[\big]{(\vI-B_{X})_{k,j}\beta_{\tCH}^{k}}\cdot\frac{2\var(\epsilon^0)}{\var(\epsilon^k)}=0.
  \end{equation*}
  This completes the proof of Corollary~\ref{thm:optimal_OLS}.
\end{proof}

\pagebreak

\section{Proposed default parameters}\label{sec:defaults}

We only consider the case of linear regression. This means that we
assume that employed regression procedure is ordinary least
squares. For the stability and prediction score we suggest the
following.
\begin{itemize}
\item \textbf{Stability score:} Depending on the number of
  environments either use a Chow test if there are only very few
  environments (less than or equal to $3$), otherwise use a scaled
  residual test based on an appropriate test statistic as discussed in
  Section~\ref{sec:stabscore}.
\item \textbf{Prediction score:} Use the negative means squared
  prediction error together with the proposed bootstrap procedure from
  Section~\ref{sec:predscore}. If one is also interested in finding
  the non-stable blanket or the Markov blanket use the negative
  environment-wise mean squared prediction error, for the predictive
  version of stabilized regression (SRpred).
\end{itemize}
When choosing default screening settings, we suggest to distinguish two
cases: (1) A setting in which we are interested in a sparse predictive
model and (2) a setting in which we want to perform an exploratory
variable selection analysis. In setting (1):
\begin{itemize}
\item \textbf{Pre-screening:} Use $\ell^1$-type screening and screen
  to as many variables as is feasible in reasonable computational
  time.
\item \textbf{Sub-sampling:} Do not sub-sample and go over all
  subsets.
\end{itemize}
In setting (2):
\begin{itemize}
\item \textbf{Pre-screening:} Use correlation screening and screen to
  approximately $\min_e \frac{n_e}{2}$ number of variables.
\item \textbf{Sub-sampling:} Set a fixed maximal set size, that allows
  for an accurate OLS fit given the sample size. Then, randomly draw
  sets with at most this maximal size. Sample as many sets as is
  feasible computationally.
\end{itemize}
For the cutoff parameters $\alpha_{\text{stab}}\in (0.01, 0.1)$ and
$\alpha_{\text{pred}}=0.01$, has worked well empirically, but can be
adjusted depending on the setting.

\newpage

\section{Gene annotations for cholesterol biosynthesis
  example}\label{sec:bio_annotations}

Details about the relation of labeled genes in
Figure~\ref{fig:pathway_example} to the Cholesterol Biosynthesis
pathway. Genes have been grouped into 4 categories: canoncial,
canoncial adjacent pathway member, semi-evident relationship and no
clear relationship. Some justification for these choices has also been
added. Canonical pathway membership was determined based on the
existence of a gene within the target pathway in GSEA
\citep{Subramanian2005}. For determining the relationship of genes
that were not in the canonical pathway, functions were examined on
both Uniprot \citep{magrane2011} and GeneCards \citep{stelzer2016} to
determine the literature relationship between the observed gene and
the canonical target pathway.

\begin{mdframed}
  \begin{compactitem}
  \item \textbf{Cyp51} \emph{Canonical.}
  \item \textbf{Dhcr7} \emph{Canonical.}
  \item \textbf{Fdft1} \emph{Canonical.}
  \item \textbf{Fdps} \emph{Canonical.}
  \item \textbf{Hsd17b7} \emph{Canonical.}
  \item \textbf{Idi1} \emph{Canonical.}
  \item \textbf{Nsdhl} \emph{Canonical.}
  \item \textbf{Sc4mol} \emph{Canonical.}
  \item \textbf{Sqle} \emph{Canonical.}
  \item \textbf{Pmvk} \emph{Canonical.}
  \item \textbf{Gstm5} \emph{No clear relationship.} CYP450 and glutathione
    gene (drug metabolism / detoxification).
  \item \textbf{Rdh11} \emph{Canonical adjacent pathway member.}
    Unclear relationship, but part of SREBF Pathway
    (HORTON\_SREBF\_TARGETS).
  \item \textbf{Acss2} \emph{Canonical adjacent pathway member.} Acyl-CoA
    transfer protein used for sterol synthesis, part of
    HALLMARK\_CHOLESTEROL\_HOMEOSTASIS.
  \item \textbf{Mavs} \emph{Semi-evident relationship.} Potentially related
    to cholesterol through IRF3 signalling and related to SREBP2. Quite distant though.
  \item \textbf{0610007P14Rik} \emph{Semi-evident relationship.} Likely
    part of adjacent pathway. It is also known as Erg28 (ergosterol
    biosynthesis protein). Not studied very well though and mouse
    function unclear. It would be immediately downstream of
    cholesterol biosynthesis and has a pretty clear functional
    connection.
  \item \textbf{Ceacam1} \emph{No clear relationship.} Glycoprotein and
    extracellular matrix / cellular structure component
  \item \textbf{Nnt} \emph{No clear relationship.}
    Glutathione-related, and is often a molecular hit identified in
    many BXD studies due to a major functional variant that
    spontaneously occurred in the parental population and segregates
    in the population \citep{aston-mourney2007, wu2014}.
  \item \textbf{Acsl3} \emph{Canonical adjacent pathway member.} Fatty acid
    activator gene, for degradation or synthesis.
  \item \textbf{Gpx4} \emph{Semi-evident relationship.} It is a very much
    downstream gene to lipid peroxidation.
  \end{compactitem}
\end{mdframed}

\newpage

\section{Details on pathways}\label{sec:pathway_details}

We selected $7$ pathways from the Reactome Pathway Knowledgebase
\citep{Reactome} and the KEGG database \citep{KEGG}. The pathways were
selected such that they allowed for reasonable recovery (at least one
comparison method was able to get a mean pAUC of at least
$0.1$). Depending on the data type, we sub-selected the pathway genes
that were measured in the data at hand. The exact list of genes used
for each data type is listed below.
\begin{itemize}
\item \textbf{Cholesterol Biosynthesis}\\
  \emph{name:} REACTOME\_CHOLESTEROL\_BIOSYNTHESIS\\
  \emph{mRNA:} Cyp51, Dhcr24, Dhcr7, Ebp, Fdft1, Fdps, Hmgcs1,
  Hsd17b7, Idi1, Lbr, Lss, Msmo1, Mvk, Nsdhl, Pmvk, Sqle, Tm7sf2\\
  \emph{protein:} Cyp51, Dhcr24, Dhcr7, Ebp, Fdft1, Fdps, Hmgcs1,
  Hsd17b7, Idi1, Lbr, Lss, Mvk, Nsdhl, Pmvk, Sc4mol, Sqle, Tm7sf2
\item \textbf{Citric Acid Cycle TCA Cycle}\\
  \emph{name:} REACTOME\_CITRIC\_ACID\_CYCLE\_TCA\_CYCLE\\
  \emph{mRNA:} Aco2, Cs, Dld, Dlst, Fh1, Idh2, Idh3a, Idh3g, Mdh2,
  Ogdh, Sdha, Sdhb, Sdhc, Sdhd, Sucla2, Suclg1, Suclg2\\
  \emph{protein:} Aco2, Cs, Dld, Dlst, Fh1, Idh2, Idh3a, Idh3g, Mdh2,
  Ogdh, Sdha, Sdhb, Sdhc, Sdhd, Sucla2, Suclg1, Suclg2
\item \textbf{Respiratory Electron Transport}\\
  \emph{name:} REACTOME\_RESPIRATORY\_ELECTRON\_TRANSPORT:\\
  \emph{mRNA:} Cox4i1, Cox5a, Cox5b, Cox6a1, Cox6b1, Cox6c, Cox7a2l,
  Cox7b, Cox7c, Cyc1, Cycs, Etfa, Etfb, Etfdh, Ndufa1, Ndufa10,
  Ndufa11, Ndufa12, Ndufa13, Ndufa2, Ndufa3, Ndufa4, Ndufa5, Ndufa6,
  Ndufa7, Ndufa8, Ndufa9, Ndufab1, Ndufb10, Ndufb2, Ndufb3, Ndufb4,
  Ndufb5, Ndufb6, Ndufb7, Ndufb8, Ndufb9, Ndufc2, Ndufs1, Ndufs2,
  Ndufs3, Ndufs4, Ndufs5, Ndufs6, Ndufs7, Ndufs8, Ndufv1, Ndufv2,
  Ndufv3, Sdha, Sdhb, Sdhc, Sdhd, Uqcr11, Uqcrb, Uqcrc1, Uqcrc2,
  Uqcrfs1, Uqcrh, Uqcrq\\
  \emph{protein:} Cox4i1, Cox5a, Cox5b, Cox6a1, Cox6b1, Cox6c,
  Cox7a2l, Cox7b, Cox7c, Cyc1, Cycs, Etfa, Etfb, Etfdh, Ndufa1,
  Ndufa10, Ndufa11, Ndufa12, Ndufa13, Ndufa2, Ndufa3, Ndufa4, Ndufa5,
  Ndufa6, Ndufa7, Ndufa8, Ndufa9, Ndufab1, Ndufb10, Ndufb2, Ndufb3,
  Ndufb4, Ndufb5, Ndufb6, Ndufb7, Ndufb8, Ndufb9, Ndufc2, Ndufs1,
  Ndufs2, Ndufs3, Ndufs4, Ndufs5, Ndufs6, Ndufs7, Ndufs8, Ndufv1,
  Ndufv2, Ndufv3, Sdha, Sdhb, Sdhc, Sdhd, Uqcr11, Uqcrb, Uqcrc1,
  Uqcrc2, Uqcrfs1, Uqcrh, Uqcrq
\item \textbf{Ribosome}\\
  \emph{name:} KEGG\_RIBOSOME\\
  \emph{mRNA:} Fau, Rpl10, Rpl10a, Rpl11, Rpl12, Rpl13, Rpl13a, Rpl14,
  Rpl15, Rpl17, Rpl18, Rpl18a, Rpl19, Rpl21, Rpl22, Rpl22l1, Rpl23,
  Rpl24, Rpl26, Rpl27, Rpl28, Rpl29, Rpl3, Rpl30, Rpl31, Rpl32, Rpl34,
  Rpl35, Rpl35a, Rpl36, Rpl36a, Rpl37, Rpl37a, Rpl38, Rpl39, Rpl4,
  Rpl5, Rpl6, Rpl7, Rpl7a, Rpl8, Rpl9, Rplp0, Rplp1, Rplp2, Rps10,
  Rps11, Rps12, Rps13, Rps15, Rps15a, Rps16, Rps17, Rps18, Rps19,
  Rps2, Rps20, Rps21, Rps23, Rps24, Rps25, Rps26, Rps27, Rps27a,
  Rps27l, Rps28, Rps29, Rps3, Rps4x, Rps5, Rps6, Rps7, Rps8, Rps9,
  Rpsa\\
  \emph{protein:} Fau, Rpl10, Rpl10a, Rpl11, Rpl12, Rpl13, Rpl13a,
  Rpl14, Rpl15, Rpl17, Rpl18, Rpl18a, Rpl19, Rpl21, Rpl22, Rpl22l1,
  Rpl23, Rpl24, Rpl26, Rpl27, Rpl28, Rpl29, Rpl3, Rpl30, Rpl31, Rpl32,
  Rpl34, Rpl35, Rpl35a, Rpl36, Rpl36a, Rpl37, Rpl37a, Rpl38, Rpl39,
  Rpl4, Rpl5, Rpl6, Rpl7, Rpl7a, Rpl8, Rpl9, Rplp0, Rplp1, Rplp2,
  Rps10, Rps11, Rps12, Rps13, Rps15, Rps15a, Rps16, Rps17, Rps18,
  Rps19, Rps2, Rps20, Rps21, Rps23, Rps24, Rps25, Rps26, Rps27,
  Rps27a, Rps27l, Rps28, Rps29, Rps3, Rps4x, Rps5, Rps6, Rps7, Rps8,
  Rps9, Rpsa
\item \textbf{Mitochondrial Ribosome}\\
  \emph{name:} REACTOME\_MITOCHONDRIAL\_TRANSLATION\\
  \emph{mRNA:} Mrpl1, Mrpl10, Mrpl11, Mrpl12, Mrpl13, Mrpl15, Mrpl16,
  Mrpl17, Mrpl18, Mrpl19, Mrpl2, Mrpl21, Mrpl22, Mrpl23, Mrpl24,
  Mrpl27, Mrpl28, Mrpl3, Mrpl34, Mrpl37, Mrpl38, Mrpl39, Mrpl4,
  Mrpl40, Mrpl41, Mrpl43, Mrpl45, Mrpl46, Mrpl47, Mrpl49, Mrpl50,
  Mrpl53, Mrpl55, Mrpl9, Mrps11, Mrps15, Mrps16, Mrps17, Mrps18b,
  Mrps2, Mrps21, Mrps22, Mrps23, Mrps24, Mrps25, Mrps26, Mrps27,
  Mrps28, Mrps30, Mrps31, Mrps33, Mrps34, Mrps35, Mrps36, Mrps5,
  Mrps6, Mrps7, Mrps9\\
  \emph{protein:} Mrpl1, Mrpl10, Mrpl11, Mrpl12, Mrpl13, Mrpl15,
  Mrpl16, Mrpl17, Mrpl18, Mrpl19, Mrpl2, Mrpl21, Mrpl22, Mrpl23,
  Mrpl24, Mrpl27, Mrpl28, Mrpl3, Mrpl34, Mrpl37, Mrpl38, Mrpl39,
  Mrpl4, Mrpl40, Mrpl41, Mrpl43, Mrpl45, Mrpl46, Mrpl47, Mrpl49,
  Mrpl50, Mrpl53, Mrpl55, Mrpl9, Mrps11, Mrps15, Mrps16, Mrps17,
  Mrps18b, Mrps2, Mrps21, Mrps22, Mrps23, Mrps24, Mrps25, Mrps26,
  Mrps27, Mrps28, Mrps30, Mrps31, Mrps33, Mrps34, Mrps35, Mrps36,
  Mrps5, Mrps6, Mrps7, Mrps9
\item \textbf{Proteasome}\\
  \emph{name:} KEGG\_PROTEASOME\\
  \emph{mRNA:} Psma1, Psma2, Psma3, Psma4, Psma5, Psma6, Psma7, Psma8,
  Psmb1, Psmb10, Psmb2, Psmb3, Psmb4, Psmb5, Psmb6, Psmb7, Psmb8,
  Psmb9, Psmc1, Psmc2, Psmc3, Psmc4, Psmc5, Psmc6, Psmd1, Psmd11,
  Psmd12, Psmd13, Psmd14, Psmd2, Psmd3, Psmd4, Psmd6, Psmd7, Psmd8,
  Psme1, Psme2, Psme3\\
  \emph{protein:} Psma1, Psma2, Psma3, Psma4, Psma5, Psma6, Psma7,
  Psma8, Psmb1, Psmb10, Psmb2, Psmb3, Psmb4, Psmb5, Psmb6, Psmb7,
  Psmb8, Psmb9, Psmc1, Psmc2, Psmc3, Psmc4, Psmc5, Psmc6, Psmd1,
  Psmd11, Psmd12, Psmd13, Psmd14, Psmd2, Psmd3, Psmd4, Psmd6, Psmd7,
  Psmd8, Psme1, Psme2, Psme3
\item \textbf{Oxphos NoATPase}\\
  \emph{name:} KEGG\_OXPHOS\_NoATPase\\
  \emph{mRNA:} Atp5a1, Atp5b, Atp5c1, Atp5d, Atp5e, Atp5f1, Atp5h,
  Atp5j, Atp5j2, Atp5l, Atp5o, Cox17, Cox4i1, Cox5a, Cox5b, Cox6a1,
  Cox6a2, Cox6b1, Cox6c, Cox7a1, Cox7a2, Cox7a2l, Cox7b, Cox7c, Cyc1,
  Ndufa1, Ndufa10, Ndufa11, Ndufa2, Ndufa3, Ndufa4, Ndufa5, Ndufa6,
  Ndufa7, Ndufa8, Ndufa9, Ndufab1, Ndufb10, Ndufb2, Ndufb3, Ndufb4,
  Ndufb5, Ndufb6, Ndufb7, Ndufb8, Ndufb9, Ndufc2, Ndufs1, Ndufs2,
  Ndufs3, Ndufs4, Ndufs5, Ndufs6, Ndufs7, Ndufs8, Ndufv1, Ndufv2,
  Ndufv3, Sdha, Sdhb, Sdhc, Sdhd, Uqcr10, Uqcr11, Uqcrb, Uqcrc1,
  Uqcrc2, Uqcrfs1, Uqcrh, Uqcrq\\
  \emph{protein:} Atp5a1, Atp5b, Atp5c1, Atp5d, Atp5e, Atp5f1, Atp5h,
  Atp5j, Atp5j2, Atp5l, Atp5o, ATP8, Cox17, Cox4i1, Cox5a, Cox5b,
  Cox6a1, Cox6a2, Cox6b1, Cox6c, Cox7a1, Cox7a2, Cox7a2l, Cox7b,
  Cox7c, Cyc1, CYTB, ND1, ND3, ND5, Ndufa1, Ndufa10, Ndufa11, Ndufa2,
  Ndufa3, Ndufa4, Ndufa5, Ndufa6, Ndufa7, Ndufa8, Ndufa9, Ndufab1,
  Ndufb10, Ndufb2, Ndufb3, Ndufb4, Ndufb5, Ndufb6, Ndufb7, Ndufb8,
  Ndufb9, Ndufc2, Ndufs1, Ndufs2, Ndufs3, Ndufs4, Ndufs5, Ndufs6,
  Ndufs7, Ndufs8, Ndufv1, Ndufv2, Ndufv3, Sdha, Sdhb, Sdhc, Sdhd,
  Uqcr10, Uqcr11, Uqcrb, Uqcrc1, Uqcrc2, Uqcrfs1, Uqcrh, Uqcrq
\end{itemize}

\newpage

\section{Biological pathway analysis (all
  results)}\label{sec:bio_results}

\subsection{mRNA}

\begin{figure}[h]
  \centering
  \resizebox{\textwidth}{!}{
    \includegraphics[width=\textwidth]{plots/mouse/KEGG_RIBOSOME_diet_mRNA.pdf}}
  \resizebox{\textwidth}{!}{
    \includegraphics[width=\textwidth]{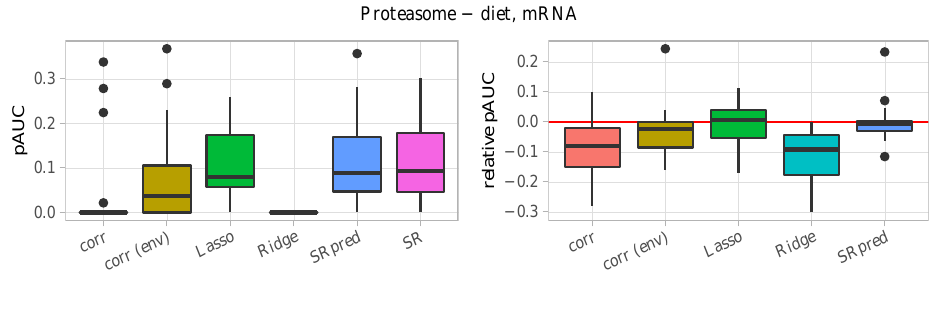}}
  \resizebox{\textwidth}{!}{
    \includegraphics[width=\textwidth]{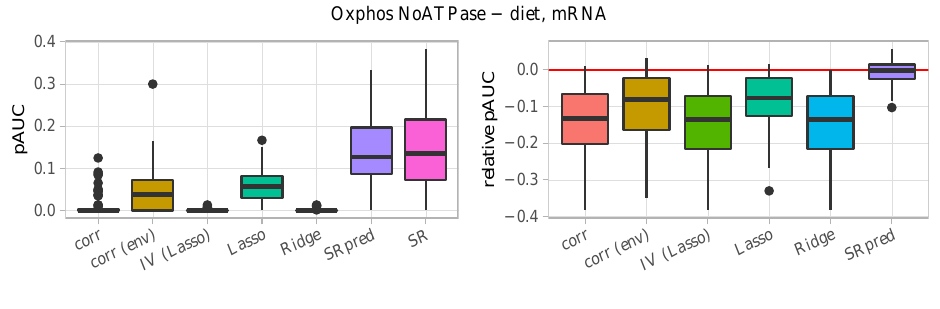}}
  \caption{First part of results for recovery analysis from
    Section~\ref{sec:recovery} using mRNA data. Analogous to
    Figure~\ref{fig:ribo_pathway_recovery}.}
  \label{fit:pathway_mRNA1}
\end{figure}

\begin{figure}[h]
  \centering
  \resizebox{\textwidth}{!}{
    \includegraphics[width=\textwidth]{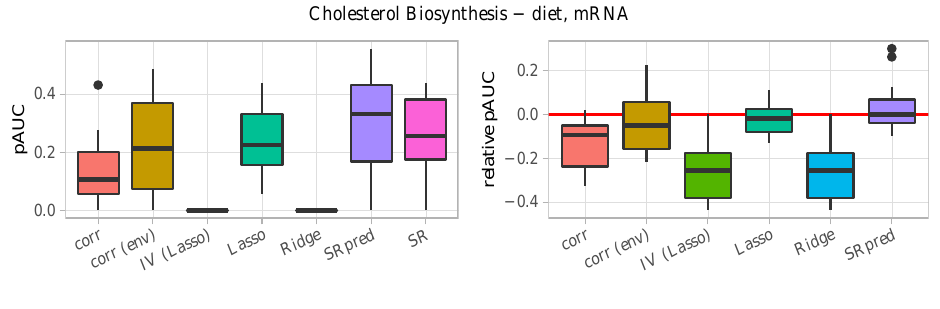}}
  \resizebox{\textwidth}{!}{
    \includegraphics[width=\textwidth]{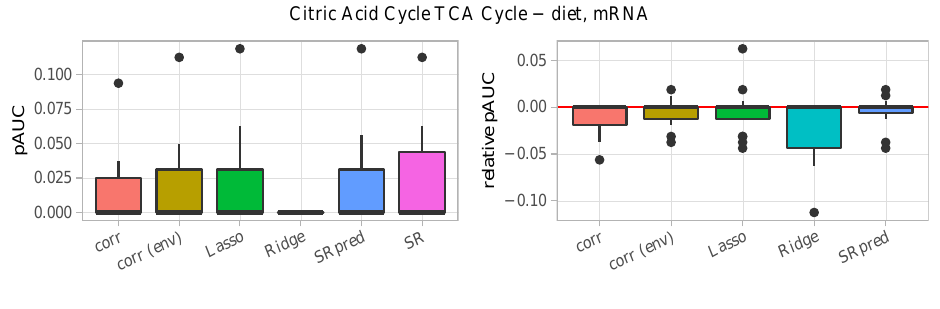}}
  \resizebox{\textwidth}{!}{
    \includegraphics[width=\textwidth]{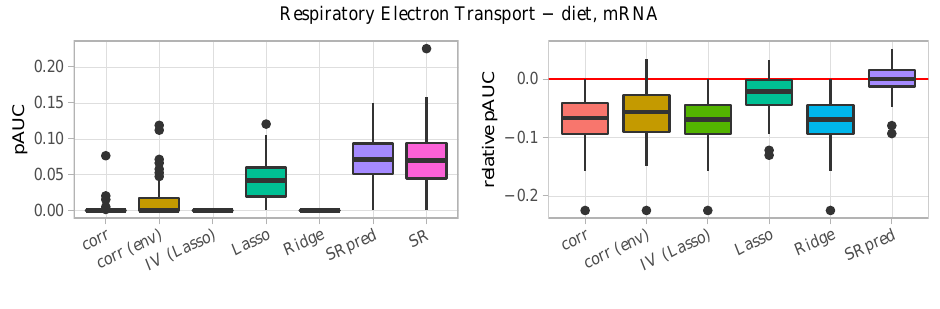}}
  \resizebox{\textwidth}{!}{
    \includegraphics[width=\textwidth]{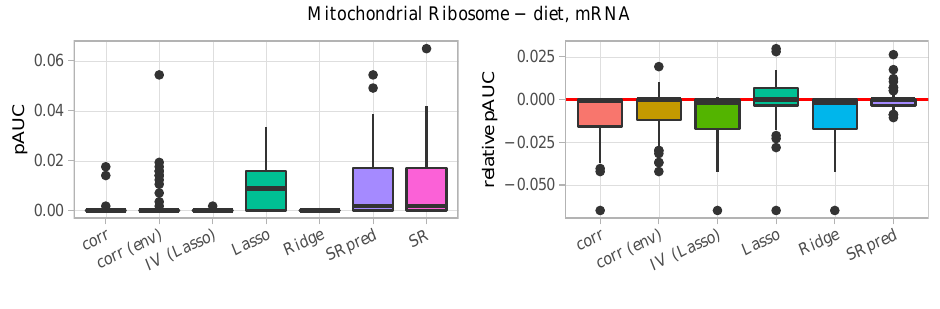}}
  \caption{Second part of results for recovery analysis from
    Section~\ref{sec:recovery} using mRNA data. Analogous to
    Figure~\ref{fig:ribo_pathway_recovery}.}
  \label{fit:pathway_mRNA2}
\end{figure}

\FloatBarrier
\newpage

\subsection{Protein}

\begin{figure}[h]
  \centering
  \resizebox{\textwidth}{!}{
    \includegraphics[width=\textwidth]{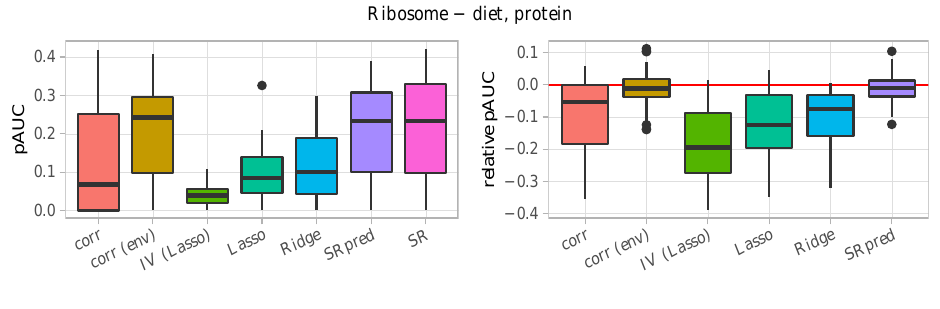}}
  \resizebox{\textwidth}{!}{
    \includegraphics[width=\textwidth]{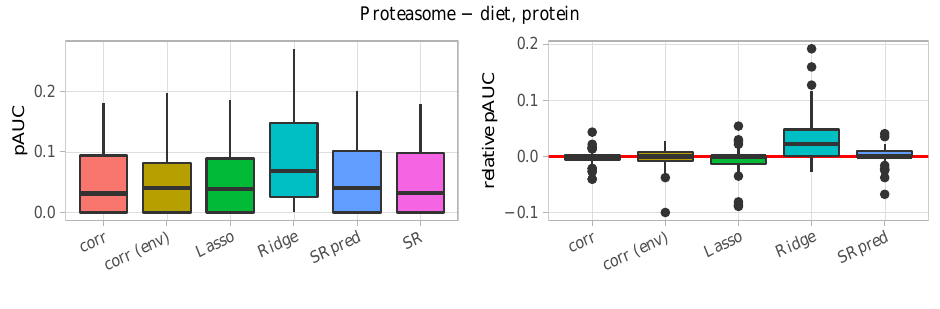}}
  \resizebox{\textwidth}{!}{
    \includegraphics[width=\textwidth]{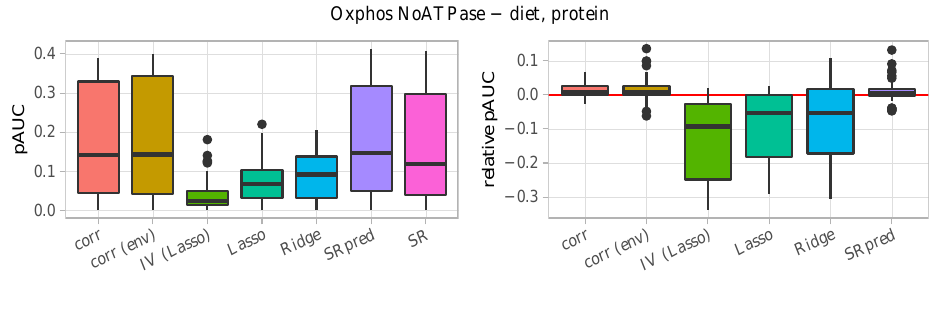}}
  \caption{First part of results for recovery analysis from
    Section~\ref{sec:recovery} using protein data. Analogous to
    Figure~\ref{fig:ribo_pathway_recovery}.}
  \label{fit:pathway_Prot1}
\end{figure}

\begin{figure}[h]
  \centering
  \resizebox{\textwidth}{!}{
    \includegraphics[width=\textwidth]{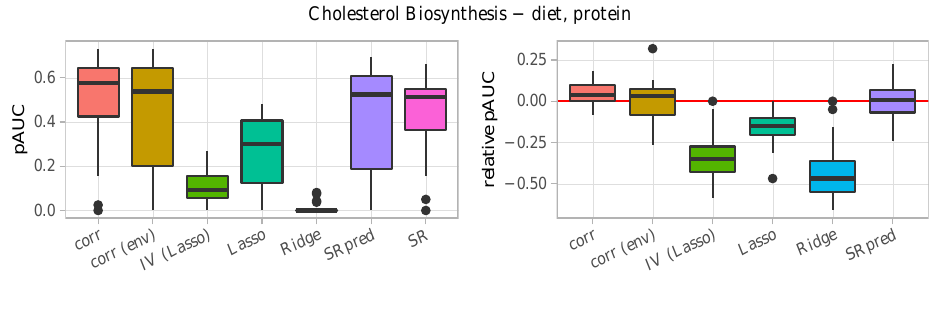}}
  \resizebox{\textwidth}{!}{
    \includegraphics[width=\textwidth]{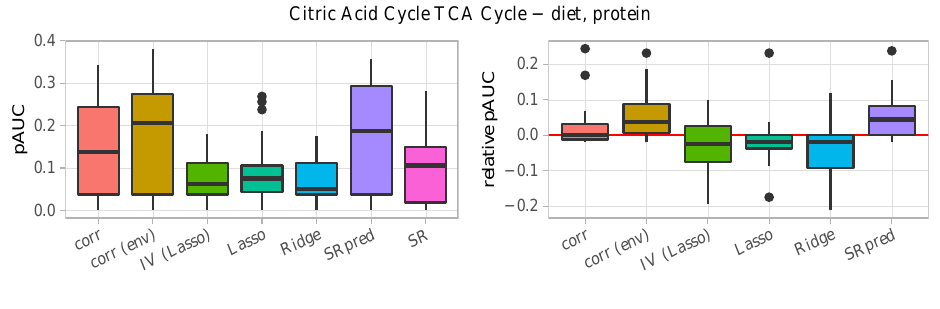}}
  \resizebox{\textwidth}{!}{
    \includegraphics[width=\textwidth]{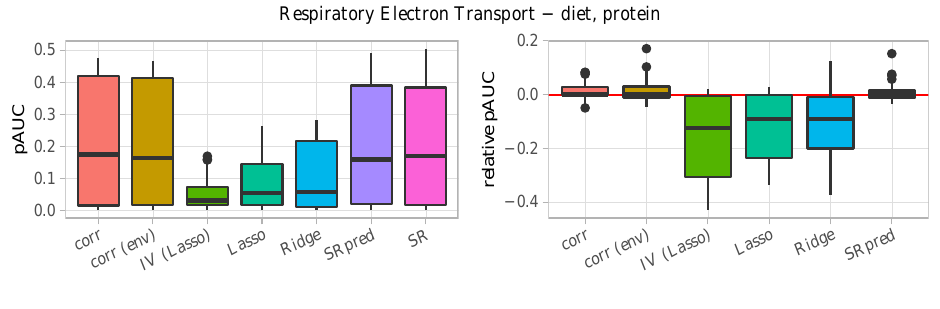}}
  \resizebox{\textwidth}{!}{
    \includegraphics[width=\textwidth]{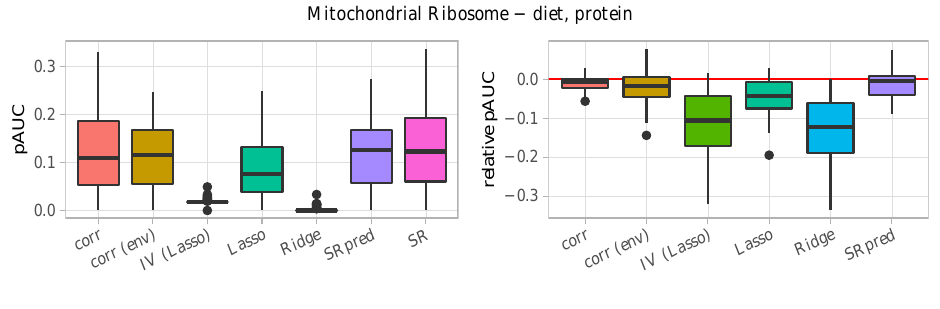}}
  \caption{Second part of results for recovery analysis from
    Section~\ref{sec:recovery} using protein data. Analogous to
    Figure~\ref{fig:ribo_pathway_recovery}.}
  \label{fit:pathway_Prot2}
\end{figure}

\FloatBarrier

%%% Local Variables:
%%% mode: latex
%%% TeX-master: "StabilizedRegression"
%%% End:

\end{document}